\documentclass[smallextended]{svjour3}       
\smartqed  
\usepackage{graphicx}
\begin{document}

\title{Multiple populations in globular clusters
}
\subtitle{Lessons learned from the Milky Way globular clusters}


\author{Raffaele G. Gratton \and
        Eugenio Carretta \and
        Angela Bragaglia }

\authorrunning{R.G. Gratton et al.} 

\institute{R.G. Gratton \at
              INAF - Osservatorio Astronomico di Padova \\
              Phone: +39-049-8293442\\
              Fax: +39-049-8759840\\
              \email{raffaele.gratton@oapd.inaf.it}           
           \and
           E. Carretta \at
              INAF - Osservatorio Astronomico di Bologna \\
              Phone: +39-051-2095776\\
              \email{eugenio carretta@oabo.inaf.it}           
           \and
           A. Bragaglia \at
              INAF - Osservatorio Astronomico di Bologna \\
              Phone: +39-051-2095770\\
              \email{angela.bragaglia@oabo.inaf.it}           
}

\date{Received: date / Accepted: date}

\maketitle

\begin{abstract}
Recent progress in studies of globular clusters has shown that they
are not simple stellar populations, being rather made of multiple generations.
Evidence stems both from photometry and spectroscopy. A new paradigm is
then arising for the formation of massive star clusters, which includes
several episodes of star formation. While this provides an explanation for several
features of globular clusters, including the second parameter problem, it
also opens new perspectives about the relation between globular clusters
and the halo of our Galaxy, and by extension of all populations with a
high specific frequency of globular clusters, such as, e.g., giant elliptical
galaxies. We review progress in this area, focusing on the most
recent studies. Several points remain to be properly understood, in particular
those concerning the nature of the polluters producing the abundance
pattern in the clusters and the typical timescale,
the range of cluster masses where this phenomenon is active, and the relation
between globular clusters and other satellites of our Galaxy.

\end{abstract}

\section{Introduction \label{intro}} 

A single stellar population is a set of stars having the same age and initial
chemical composition (in terms of both He and metal abundances), and different
mass, distributed according to an initial mass function (IMF).  This definition,
which is described by a single isochrone in the colour-magnitude  diagram, may
possibly be extended to include binaries. Single stellar population is a tool 
widely used in stellar and galactic evolution context. Stellar populations in 
galaxies are usually assumed to be reproduced by suitably weighted sums of
single stellar populations. Stellar clusters were usually considered good
examples of single stellar populations. Many concepts of  stellar evolution were
developed in order to describe the location of stars in  the colour-magnitude
diagram of clusters using the single stellar population concept. A short list
includes the nature of  blue stragglers, the inclusion of mass loss in stellar
models, the  explanation of the horizontal branch (HB) including
semi-convection, most  aspects of the asymptotic giant branch (AGB) evolution,
the evolution of stars  along the white dwarf cooling sequence, the
transformations between  theoretical and observational planes, etc. Single
stellar population is clearly a very useful concept. As many other
approximations, it however fails when subject to deeper scrutiny. The
colour-magnitude diagram of a globular cluster may look  as an isochrone at
first glance,  that is, a globular cluster appears close to a single stellar
population; however, a closer examination soon reveals  for instance, that the
HB of most globular clusters cannot be reproduced by a single stellar
population. In a way,  the fact that real clusters are not exactly single
stellar populations should be obvious, since  instantaneous star formation and
complete chemical homogeneity of the original molecular  cloud are not to be
expected. However, the departures of globular clusters from single stellar
population are much  larger and significant than simply a small spread of the
formation epoch or  original inhomogeneities. It can be shown that galaxies
cannot be reproduced by  a suitably weighted sum of a number of single stellar
population-globular clusters. Rather, multiple populations (MPs)  in globular
clusters provide information on how they formed, and on the likely relation
with  the host galaxy. Throughout this review, we will use the term ``multiple
populations" as the synonymous of ``multiple generations of stars" that can be
distinguished either from their spectra or from multiple sequences in the
colour-magnitude diagram. Other authors prefer limiting this definition only to
those cases where multiple main sequences or subgiant branches - but oddly not
horizontal branches or red giant branches (RGBs) - have been observed. We think
that such a distinction only arises for historical reasons and does not help to
understand the wide implications of the multiple population question.
Observation of multiple main sequences and subgiant branches require photometry
of exquisite quality, and is up to now limited to only a few cases - a sample
which is however growing steadily in the last few years. The case of NGC~6397, a
globular cluster thought to have an astonishingly narrow main sequence, which
has been shown to consists of two close but distinct sequences by Milone et al.
(2011), is emblematic. While it is clear that other confirmations of the
correlation between spectroscopic and photometric properties would be welcome,
the observational data are already sufficiently strong to make the conclusion
about its reality robust, and justify our use of the multiple population
concept.

Progress in the last few years have in fact shown that globular cluster
formation   is a complex phenomenon. Scenarios for globular cluster formation
now include several  generations of stars, that we may schematically divide into
a primordial, first generation  and a second generation, although the second
generation likely  consists of several distinct sub-populations. The second
generation stars form from material  polluted by the ejecta of a fraction of the
first generation population (Gratton et al. 2001,  2004; Ramirez \& Cohen 2002;
Carretta et al. 2010a). Evidence for the presence of these different 
populations stems from their chemical composition, in particular the abundances 
of the light elements involved in proton-capture processes (C, N, O, F, Na, Mg,
Al,  Si: Osborn 1971; Cottrell and Norris 1981; Kraft 1978, 1994; Gratton et al.
2001, 2004; see Carretta et al. 2009a, 2009b, and references therein), as well
as  from the splitting of sequences in the colour-magnitude diagram (see Piotto 
2009). Most of the stars currently in globular clusters belong to the second
generation that is characterised  by a peculiar composition, rarely found among
field stars. The residual  component of the first generation, low-mass stars
with the composition typical of  the ejecta of core-collapse supernovae (e.g.
Truran \& Arnett 1971) and similar to that of field halo stars, is still present
in globular clusters, but is only a  third of the current total population.
Since the peculiar composition of the second generation stars requires that they
formed from the ejecta of only a fraction of the first generation stars  (the so
called ``polluters"), it may be concluded that globular clusters were created
within massive episodes of star formation and that most of their original
population has  been lost. This has renewed the interest for the relation
existing between globular clusters  and field halo: it is even possible that
most of the halo of our Galaxy - and  by extension of other environments with a
similar high specific frequency of globular clusters - formed within these
massive star formation episodes.

Many aspects of this scenario are still unclear and currently debated. They
concern the timescale of the phenomenon and the related issue of the nature of
the polluters; the need for ``dilution" of the processed  material with
unprocessed material (see Prantzos \& Charbonnel 2006) and the origin  of the
diluting material; the construction of realistic hydrodynamical models; the
identification and characterisation of young, present-day counterparts;    the
connection between different observational aspects of the phenomenon (photometry
vs  spectroscopy; the evidence in different evolutionary phases; the  connection
with  anomalies in the colour-magnitude diagram such as extended, blue HBs; the
AGB; etc.); the range of masses of the original population and the relation with
open clusters, nuclear star clusters, ultra-compact dwarf galaxies, and dwarf
spheroidals (dSph); the r\^ole  of globular clusters in galactic evolution,
etc.  

The reader should also be alerted that in addition to the scenario we discuss in
this review, there are several mechanisms that can potentially produce chemical 
inhomogeneities in globular clusters. They include:
\begin{itemize}
\item Original inhomogeneities in the material from which the globular cluster
formed (see e.g. Kraft 1979). At variance with earlier findings, star forming
regions and very young clusters are now usually found to be chemically
homogeneous in  spite of evidence of rather prolonged star formation; this
suggests that  interstellar matter is effectively well mixed and that
self-enrichment  was not relevant (see e.g., Shen et al. 2005; D'Orazi \&
Randich 2009;  Nieva \& Simon-Diaz 2011, and references therein). However,
globular clusters are much more  massive objects, and likely formed in very
large although compact star  forming regions. Chemical inhomogeneities might
either be primordial, caused by  inadequate mixing of the interstellar matter
from which the stars formed, or through some -  perhaps limited - capability of
retaining ejecta of the most rapidly evolving  stars, since the formation of the
globular cluster must have taken some time. Various authors  (see the review by
McKee \& Ostriker 2007 and references therein) suggest that  star formation
within a giant molecular cloud occurs on timescales  comparable to a few times
(1-3) the crossing time (roughly, $t_{\rm cr}  \sim 10~M_6^{1/4}$~Myr, where
$M_6$\ is the mass of the cloud in units of  $10^6$~M$_\odot$). Keeping in mind
that the final cluster is a few percent of  the original giant molecular cloud
(see below), a timescale of 30~Myr for the formation of the  primordial
population of a cluster such as M~54 and $\omega$~Cen (currently a few 
$10^6$~M$_\odot$) seems fully reasonable. Clusters smaller in size by an order
of  magnitude (that is, typical clusters with $M_V \sim -7.8$) should have
required  a shorter but not negligible time ($\sim 10-20$~Myr). On the other
hand,  Galactic giant molecular clouds seem to have an upper limit to their mass
of a few $10^6$~M$_\odot$;  larger masses are usually called Giant Molecular
Associations, because they  fragment in smaller pieces. It is not clear whether
this depends on metallicity or  environment. This is likely the reason why
globular clusters do not form any more in the Milky Way, given the low
efficiency with which giant molecular clouds can be transformed in clusters.
Also,  fragmentation of the larger giant molecular clouds may lead to more
complicate geometries, and then  to smaller efficiencies in the cloud-to-cluster
transformation.
\item Peculiar evolution of individual stars, either because they are members
of  interacting binaries, or because some parameter such as, for example,
internal rotation, may  cause significant alterations to the surface chemical
composition, due e.g. to a high mass loss rate or meridional circulation
(Sweigart \& Mengel 1979). This  mechanism has long been considered in
connection to the Na-O anticorrelation (see  the reviews by Kraft 1979, 1994),
as well as the HB second parameter (see the review  by Catelan 2009).
Interacting binaries have certainly a r\^ole in explaining the  blue stragglers
and other peculiar objects, such as millisecond pulsars, found in globular
clusters,  but these are only a small minority of the cluster stars. The impact
of peculiar  evolution on the global properties of globular clusters has never
been assessed in a definitive  way in spite of many attempts. It is possible and
even likely that overall mass loss  is slightly different from star-to-star in
globular clusters (see the discussion in Catelan 2009);  however, this effect is
probably small and cannot explain most of the phenomenology  and regularities
discussed throughout this review.
\item Merging of globular clusters. While this event is very unlikely to occur
in the halo of  the Milky Way owing to the very large relative velocities of
globular clusters, it might happen in  dSph's such as, e.g. Sagittarius (Sgr),
where the spread of velocities  within the galaxy are comparable to that within
the globular clusters. Globular clusters formed in a dSph may  then be inherited
by the Milky Way when the dSph is tidally destroyed, as is  observed for Sgr. In
these conditions, dynamical friction may be very efficient,  leading the most
massive clusters to spiral toward the dSph centre and merge (see  e.g.
Bellazzini et al. 2008). A similar fate is likely to occur for binary clusters
such as those frequently observed in the Magellanic Clouds  (Bhatia \& 
Hatzidimitriou 1988).
\item Interstellar matter originally unrelated to the cluster may cumulate
around deep potential  wells and provide material for new episodes of star
formation. This mechanism is likely active around the central black hole in
galactic nuclei, and might  explain the origin of the nuclear star clusters
observed in almost all late type  galaxies including the Milky Way (B\"oker
2010) where there seems to be repeated episodes  of star formation (Krabbe et
al. 1995; Rossa et al. 2006; Walcher et al. 2006). The  potential wells of these
objects are typically much deeper than for globular clusters, except possibly
for the most massive ones. Accretion of interstellar matter by globular
clusters  has been recently modelled by various authors (Bekki \& Mackey 2009;
Conroy \&  Spergel 2011, but see D'Ercole et al. 2011 for the limits on the
amount of pristine interstellar matter that may be involved in the second
generation formation to reproduce the observed abundance pattern), and  might be
an  attractive explanation for those cases where the  relative velocities are
small,  such as the Magellanic Cloud clusters with multiple main sequence
turn-off's (TO's).
\end{itemize}
The chemical imprinting of these various mechanisms is likely  different from
that originated by the scenario of multiple populations related to the cluster 
formation we will discuss throughout this review. For instance, most of them
cannot produce O-poor stars, typically observed only in the cluster environment.
However, since they may perhaps occasionally occur, they may complicate the
observational pattern. One of the main purposes  of current investigations is to
separate the aftermaths of these different  mechanisms, in order to properly
understand what globular clusters are and how they formed.

In this review, we describe recent progress in the investigation of multiple
populations in globular clusters. We mainly focus on the progress made after the
last review by  Gratton et al. (2004), in turn an update of the review by Kraft
(1994). Our presentation is biased toward evidence from spectroscopy, which  is
the field where the authors are more expert. A recent review of evidence from
photometry can be found in  Piotto (2009). Excellent reviews on related topics
include Helmi (2008), which discusses the characteristics and origin of the
Galactic halo; Brodie \& Strader (2006), which presents evidence on
extragalactic globular cluster systems; and  Portegies-Zwaart et al. (2010) on
young massive stellar clusters. Very useful is also the series of papers on
stellar clusters which appeared in Volume 368 of the Philosophical Transactions
of the Royal Society.

The review is organised as follows. In Section~\ref{sec:2} we present the
evidence for multiple stellar populations along the colour magnitude diagram.
Both photometric and spectroscopic results will be reviewed. In
Section~\ref{sec:3} we discuss the relevant aspects of stellar evolution and
nucleosynthesis and the related scenarios for globular cluster formation and
early evolution. In Section~\ref{sec:4} we analyse the connection between
globular clusters and the halo of our Galaxy, focusing on the impact of multiple
stellar populations. A short presentation of open questions and expected
advances is given in the concluding Section.
 
\section{Observational aspects of multiple stellar populations}
\label{sec:2}

In this Section, we describe the phenomenology of the multiple populations 
along the different evolutionary sequences of globular clusters, beginning with
the  main sequence even if the first evidence came of course from the brighter, 
evolved RGB stars.

\subsection{Main Sequence (MS)\label{ms}} 

For a long time the photometric narrow evolutionary sequences were interpreted 
as proofs of the single-age, single-metallicity nature of globular clusters.
However, variations  in light elements were being found in MS stars as early as
at the beginning of  the 80's. Hesser \& Bell (1980) found different CN band
strengths in seven MS stars of 47~Tuc, similar to that found for giants;  they
noted that this implies a primordial origin, since there is no mechanism  able
to mix N to the surface in MS stars, at variance with RGB ones (see 
Sect.~\ref{rgb}). Similar findings were presented, e.g., by Cannon et al. 
(1998) together with a well developed discussion of the possible sources.  While
it was recognised that mixing was not a viable explanation because MS  low-mass
stars do not have a convective envelope able to transport the  nuclearly
processed material to the atmosphere, the other possibilities (primordial origin
caused by two or more protocluster gas clouds, self-enrichment  due to a second
generation of stars, and surface pollution caused by mass transfer  on already
formed stars) were considered and debated. The amount of data  increased, and
generally a bimodality in CN strength and an anticorrelation  between CN and CH
was found (see e.g., Cohen 1999a,b; Harbeck et al. 2003; Briley el al. 2004;
Cohen et al. 2005). More recently Kayser et al. (2008)  and Pancino et al.
(2010) have measured CN and CH in eight and 12 globular clusters,  respectively,
partly in common, explicitly interpreting their results in the  multiple
population scenario.

It was only later that measures of the other existing variations in light 
elements (Na, O, Al, Mg, see Sect. ~\ref{rgb}) could be made for stars on  the
MS. Using low resolution spectra, Briley et al. (1996) found significant 
variations in Na, correlated with CN, in six stars at the main sequence turn-off
of 47~Tuc.  The seminal work was by Gratton et al. (2001) who presented Na, O,
Mg, and Al abundances for a few main sequence turn-off stars in NGC~6752 and
NGC~6397. They analyzed UVES spectra at high resolution, finding clear
anticorrelations (especially when also unevolved,  SGB stars were considered)
between O and Na, and Mg and Al,  similar to what was present in RGB stars.
Since the temperature at which the  reactions that produce Na, and especially
those that destroy Mg (70 MK,  Arnould et al. 1999), can not be reached by
low-mass, MS stars (see  Sect.~\ref{sec:3}), this was a clear indication of an
external origin. This  was later confirmed on rather small samples by Ramirez \&
Cohen (2002, 2003)  for M~71 and M~5, Carretta et al. (2004) for 47~Tuc, Cohen
\& Melendez (2005)  for M~13, and recently by D'Orazi et al. (2010b), Lind et
al. (2011), and Monaco et al. (2011) on large samples of stars in 47~Tuc,
NGC~6397,  and M~4, where Na, O, and Li were measured.

The abundance of Li and the (possible) dispersion and (anti)correlation with  O
and Na could be robust indicators not only of mixing phenomena but of the 
nature of the polluting stars, since Li can be produced by AGB stars, but not 
by rapidly rotating massive stars (see Sect.~\ref{polluters}). Lithium in a
large  sample of MS stars has been measured only in a few, nearby globular
clusters: NGC~6397, where  Lind et al. (2009) found a Li-Na anticorrelation;
47~Tuc, for which D'Orazi et al. (2010b) found no evidence of Li-O or Li-Na
correlations; NGC~6752, where  Shen et al. (2010) measured a Li-O correlation;
and M4 for which Mucciarelli  et al. (2011) did not find a Li-O correlation,
while Monaco et al. (2011) find  some correlation, but with a very shallow
slope. 

\begin{figure}
\centering
\includegraphics[scale=0.45]{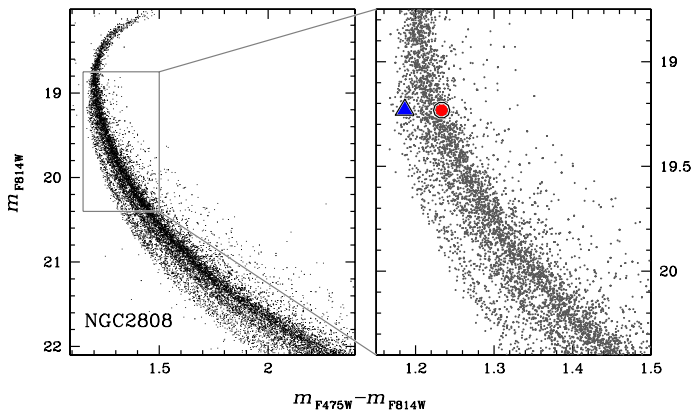}
\caption{The triple MS of NGC~2808 (Piotto et al. 2007) and the position of the
two MS stars, one on the bMS, one on the rMS, analysed by Bragaglia et al.
(2010b).}
\label{f:n2808}.
\end{figure}

Photometric evidence of multiple populations came later, but it had an immediate
impact on the  community, both because it was more easily recognisable and
because the times  were ripe for a change of perspective. Data were obtained
using the superb  capabilities of the Hubble Space Telescope (HST), which
permitted the beating down of the photometric uncertainties obtaining very clean
colour-magnitude diagrams where also subtle  features could be seen. Not
surprisingly, the first identification of multiple  MSs was in $\omega$ Cen;
Bedin et al. (2004) were able to demonstrate that it  possesses two clearly
distinct lower MSs (that became three in Villanova et al.  2007), which could
differ in He content, as suggested by Norris (2004), with the bluer sequence
enriched to Y$\sim 0.38$. Intermediate-low resolution stacked spectra of stars
on the blue and red MSs were analysed by Piotto et al. (2005), providing the
surprising result that the blue MS is about 0.3 dex more metal-rich than the red
one, which could be explained if it were enriched in He to Y$\sim 0.35-0.40$.
$\omega$~Cen had always been a peculiar case, but the same reasoning on He
enhancement was used by D'Antona et al. (2005) to explain the MS and HB of the
``normal" globular cluster NGC~2808 (see also Sect.~\ref{hb}). Soon after,
Piotto et al. (2007) presented a colour-magnitude diagram based on ACS@HST data,
showing three distinct MSs in this cluster (Fig.~\ref{f:n2808}), that could be
explained by three populations with the same age and metallicity but three
different (and discrete) He abundances. They linked the three sequences to the
three groups of stars along the Na-O anticorrelation (Carretta et al. 2006, see
Sect.~\ref{rgb}) and the three parts of the HB (Sect.~\ref{hb}) on the basis of
number counts and evolutionary considerations. The very high sensitivity of
X-shooter@VLT permitted the chemical tagging of two faint MS stars, one on the
red MS (rMS), one on the blue MS (bMS). The first has the normal composition of
a first generation star, while the latter shows depleted C and Mg, enhanced Al,
Na, and N, which is typical of second generation stars (Bragaglia et al. 2010b
and Fig.~\ref{f:n2808}); while based on a very limited sample, this appears to
be a convincing confirmation of the scenario of multiple populations in globular
clusters. Further, less spectacular cases for a spread or split MS were
presented for 47~Tuc (Anderson et al. 2009), NGC~6752 (Milone et al. 2010), and
NGC~6397 (Milone et al. 2011); di Criscienzo (2010a,b)  interpreted the MSs of
NGC~6397 and 47~Tuc in the light of a (small) spread in He  abundance.

Finally, the globular clusters in the Magellanic Clouds merit a special mention.
Bertelli et al. (2003) found an indication of a possible prolonged star
formation for the Large Magellanic Cloud cluster NGC~2173. However, the first
clear  indications of multiple population in Magellanic Cloud globular clusters
came again from HST photometry, see e.g., Mackey \& Broby Nielsen (2007); Milone
et al. (2008); Glatt et al. (2008); and Goudfrooij et al. (2011). These authors
found that many Magellanic Cloud intermediate-age clusters (age about 1-2 Gyr)
show a spread or split main sequence turn-off and often also a dual red clump;
in particular, Milone et al. examined 16 globular clusters finding a signature
of multiple main sequence turn-offs in 11 of them. If this is due to an age
difference, the implied interval is of about 200-700 Myr, even if a prolonged
star formation is a possible alternative (e.g., Rubele et al. 2010).  An
alternative solution (rotation) has been proposed by Bastian \& de Mink (2008)
and contested e.g., by Girardi et al. (2011). Finally, Keller et al. (2011)
equate the Magellanic Cloud globular clusters with extended  main sequence
turn-offs with the Galactic globular clusters and expect them to show the same
chemical peculiarities, although to a lesser extent.

\subsection{Subgiant branch (SGB) \label{sgb}} 

The SGB stars contributed, together with the even less evolved MS ones, to the 
definitive acceptance of a primordial origin for the light elements
``anomalies", since no efficient mixing is possible before the first dredge up
episode and the RGB bump (see the previous Section, and Gratton et al. 2004 for
a review of the observations and arguments).

In the case of $\omega$ Cen, the SGB had been known to present a large spread 
(e.g., Hilker \& Richtler 2000), possibly implying a spread in age. The first 
indication of an actual split was shown by Ferraro et al. (2004), who found a 
fainter SGB, connected to the highest metallicity RGB. This split was confirmed
(e.g., Bedin et al. 2004; Sollima et al. 2005) and a more complicated structure
was detected. For instance, Villanova et al. (2007) identified at least four
distinct SGBs and explained them with differences in metallicity and in age (but
the large age spread, $>4$ Gyr, is controversial) trying also to understand how
they connect to the multiple MSs and RGBs. Detailed abundance analysis, such as
the one in Pancino et al. (2011), but on larger samples of SGB stars in each of
the sub-sequences is required to settle the problem.

Impressive evidence of multiple populations comes from the very precise
photometry of an increasing number of lower-mass clusters, obtained mainly with
HST: NGC~1851 (Milone et al. 2008; in this globular cluster also ground-based
data show a split, see Zoccali et al. 2009, Milone et al. 2009), NGC~6388
(Moretti et al. 2009, combining optical HST and infrared ground-based adaptive
optic data), 47~Tuc (Anderson et al. 2009), M~22, and M~54 (Piotto et al. 2009).

NGC~1851 is the best studied of all these. As already indicated by the discovery
paper (Milone et al. 2008), the split SGB can be explained by a difference in
age ($\sim$ 1 Gyr) or in C+N+O content (He has a small effect and the required
difference in [Fe/H] would show up elsewhere in the colour-magnitude diagram).
An enhanced CNO abundance was explored e.g., by Cassisi et al. (2008) and
Ventura et al. (2009a), finding it a viable explanation. Indeed, Yong et al.
(2009) proposed a variable CNO in a small sample of RGB stars (not found by
Villanova et al. 2010, but apparently confirmed on a larger sample, Yong, still
unpublished). Also debated is the radial distribution of the two sub-branches
(e.g., Zoccali et al. 2009 and Milone et al. 2009); a more concentrated
distribution is expected for the second generation. The split, however, appears
to be present also at very large distances (Olszewski et al. 2009). A recent
discussion on this globular cluster can be found in Carretta et al. (2011a), who
prefer the merging of two globular clusters of slightly different age as an
explanation for the many unusual features of this cluster.

\subsection{Red giant branch (RGB) \label{rgb}} 

Originally, the first detections of unusual abundances (always regarding light 
elements) among cluster stars came of course from the brighter, more accessible 
RGB stars. We will not repeat here the long list of observations, starting from
Osborn (1971, who discovered two N-enhanced stars in M~10 and M~5 using DDO
photometry), passing through many low-resolution measures of CN and CH
abundances with different distributions in the various clusters, and to the
first papers that noted the existence of relations between C and N and heavier
elements (e.g, correlation between CN strength and Na, Al found by Cottrell \&
Da Costa 1981, or anticorrelation between CN and O, coupled with correlation
between CN and Na, found by Sneden et al. 1992). Neither will we insist on the
early work on the Na-O (and related Mg-Al) anticorrelations conducted mainly by
the Lick-Texas group (e.g., Kraft et al. 1993; 1997; Ivans et al. 2001; Sneden
et al. 2004, to name just a few). We refer to Kraft (1994), Gratton et al.
(2004), and Martell (2011) for more details of the studies, the interpretations
(and their evolution with growing evidence and theoretical works), and extensive
references. We wish however to recall that the old idea that the Na-O
anticorrelation can be caused by peculiar evolution of red giants in globular
clusters is untenable, as demonstrated by the fact that the same anticorrelation
is found among unevolved stars in globular clusters (see the extensive
discussion in  Gratton et al. 2004). This leaves original differences between
stars in globular clusters as the only possible solution. The only scenario
where this is possible is the multiple population scenario.

\begin{figure}
\centering
\includegraphics[scale=0.6]{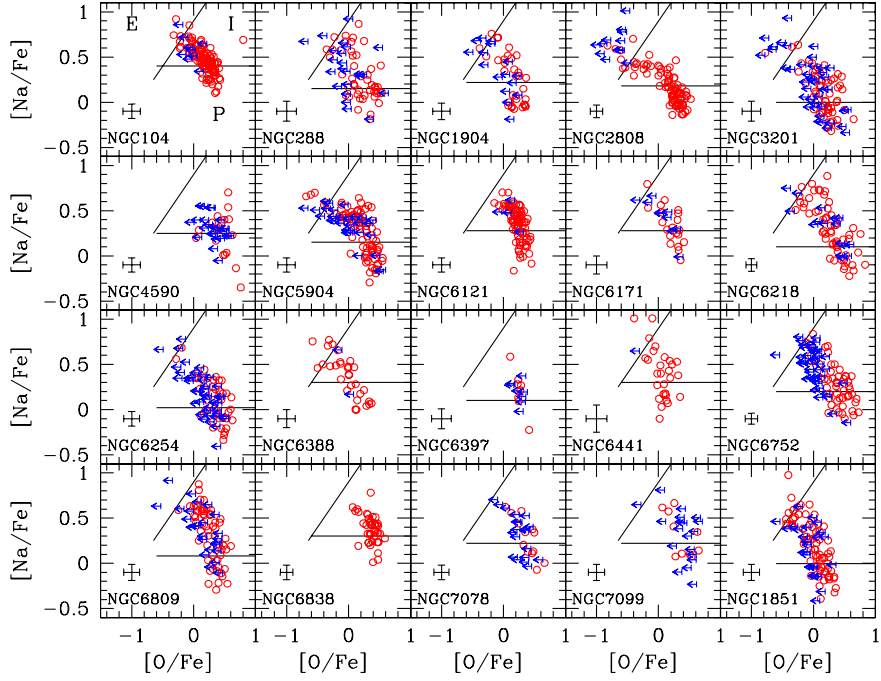} 
\caption{A modern collection of Na-O anticorrelations showing partial results of
the FLAMES survey of globular clusters (see Carretta et al. 2009a, 2010d for
full references) and demonstrating both the ubiquity of this feature and its
difference cluster-to-cluster. The lines separate the P (first generation), I,
and E (second generation) components, as defined in Carretta et al. In this
plot, red circles are measures for both O and Na, and blue arrows are measures
for Na but only upper limits for O.}
\label{f:nao}.
\end{figure}

The RGBs of many globular clusters have been recently studied using
high-resolution spectroscopy to discover intrinsic abundance variations in light
elements (C, N, O, Na, Al, Mg, Si, F) that (anti)correlate in a fashion
explained by the effects of H-burning at high temperature (see
Sect.~\ref{sec:3}). To list only works not referenced in the review by Gratton
et al. (2004), the following Milky Way clusters have been studied (in addition
to $\omega$ Cen, see below): M~3, M~13 (Sneden et al. 2004, Cohen \& Melendez
2004, Johnson et al. 2005, Yong et al. 2006); M~80 (Cavallo et al. 2004);
NGC~7492 (Cohen \& Melendez 2005); M~10, NGC~7006 (Smith et al. 2005a); NGC~2808
(Carretta et al. 2006); NGC~6218 (Johnson et al. 2006, Carretta et al. 2007b);
NGC~6752 (Carretta et al. 2007a, Yong et al. 2008); NGC~6397 (Korn et al. 2007;
Carretta et al. 2009a,b; Lind et al. 2011); NGC~6388 (Carretta et al. 2007c);
NGC~6441 (Gratton et al. 2006, 2007); NGC~6712 (Yong et al. 2008); M~4 (Smith et
al. 2005b; Marino et al. 2008; Carretta et al. 2009a,b; Monaco et al. 2011);
M~22 (Marino et al. 2009, 2011b); NGC~1851 (Yong \& Grundahl 2008; Carretta et
al. 2010d, 2011a; Villanova et al. 2010); M~53 (Martell et al. 2008); Pal 3
(Koch \& McWilliams 2009); M~5 (Carretta et al. 2009a,b; Lai et al. 2011);
NGC~288, NGC~362 (Smith et al. 2009; Carretta et al. 2009a,b); 47~Tuc, NGC~1904,
NGC~3201, NGC~4590, NGC~6171, NGC~6254, NGC~6809, NGC~6838, NGC~7078, NGC~7099
(Carretta et al. 2009a,b: these papers present results on the first 19 globular
clusters of a survey with  FLAMES@VLT, see Fig.~\ref{f:nao}); M~54 (Carretta et
al. 2010b,c); NGC~2419 (Cohen et al. 2011). In addition, similar variations have
been found in old globular clusters in the Large Magellanic Cloud (Johnson et
al. 2006; Mucciarelli et al. 2009) and Fornax (Letarte et al. 2006). 

\begin{figure}
\centering
\includegraphics[scale=0.45]{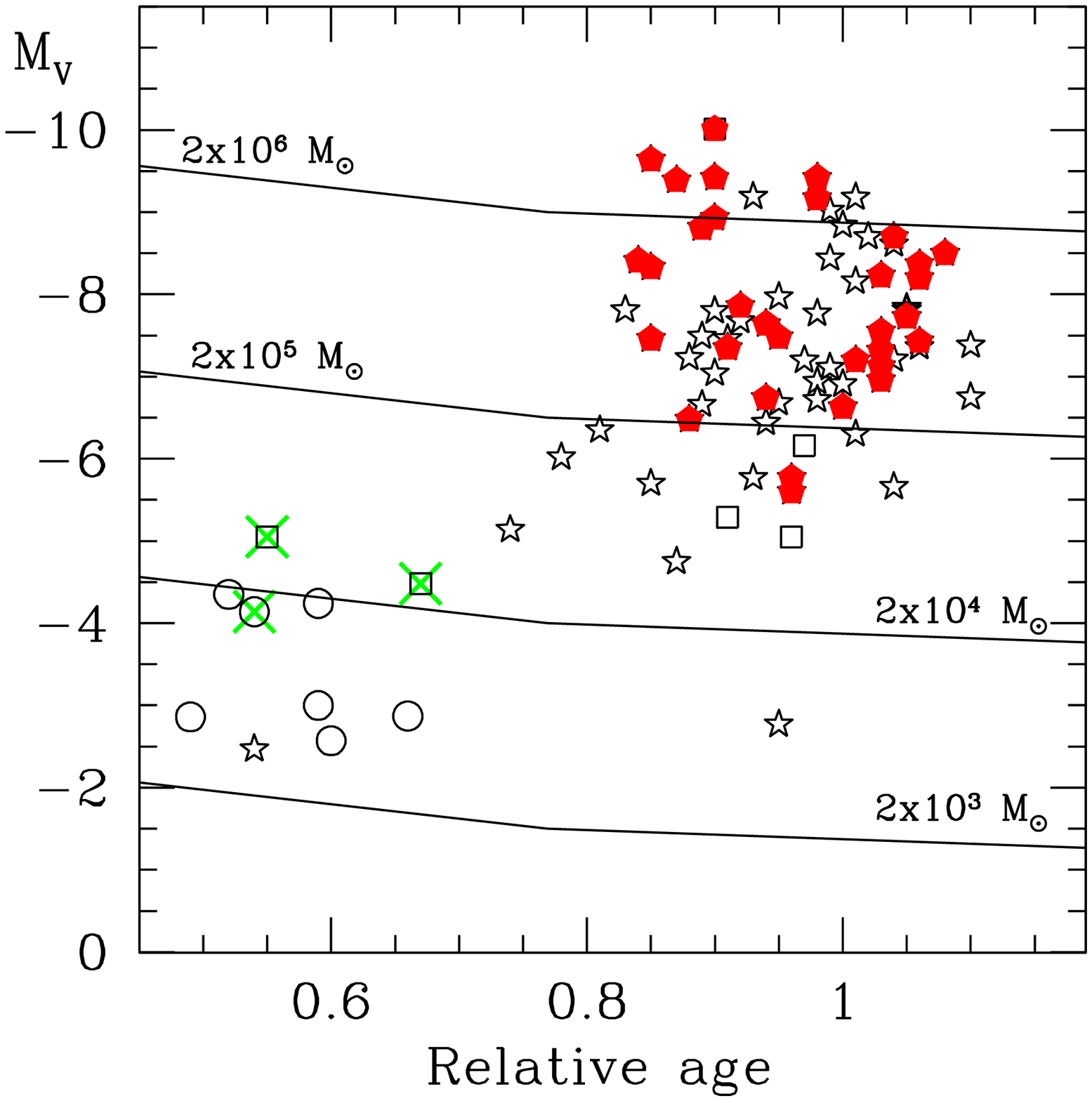}
\caption{The age-M$_V$ (a proxy for mass) distribution for clusters in the Milky
Way, the Sgr dSph, and the Large Magellanic Cloud (open squares); open clusters
are indicated by open circles; filled, red symbols indicate globular clusters
where an Na-O anticorrelation has  been found; green crosses the three clusters
where it is not confirmed (from Gratton et al. 2010b); open stars are globular
clusters not yet investigated.}
\label{f:mvage}.
\end{figure}

A (partial) misconception is that multiple populations are found only in massive
globular clusters. Multiple photometric sequences have been found first in
massive clusters ($\omega$ Cen, M~54, NGC~2808, NGC~6388, 47~Tuc) and only later
in more average globular clusters (NGC~1851, NGC~6752, M~22, or the Magellanic
Clouds ones). However, the spectroscopic evidence is that {\it all} Milky Way
globular clusters surveyed (with the possible exception of Terzan 7 and Pal 12
with only a few analysed stars: Sbordone et al. 2005, Cohen 2004) show an Na-O
anticorrelation, therefore, {\it all} of them have multiple generations of stars
(see Fig.~\ref{f:mvage}). There are however indications that the cluster mass
has important bearings on the cluster formation and evolution. For instance,
Carretta et al. (2010a) found that the extension\footnote{The  extension of the
Na-O anticorrelation can be quantitatively expressed by the  interquartile
ratio, IQR, as described e.g., in Carretta (2006).} of the Na-O anticorrelation
(i.e., the amount of polluting material, the mass of the producers, the initial
conditions, etc., see Sect. 3) increases with absolute magnitude (mass), and
this includes also M~54 and $\omega$ Cen (Carretta et al. 2010c).  Empirically,
a present-day mass larger than a few 10$^4$ M$_\odot$ seems required for a
globular cluster to show an Na-O anticorrelation (Carretta et al. 2010a). This
is further corroborated by the fact that this feature has not been found in any
open cluster, however massive (de Silva et al. 2009; Pancino et al. 2010;
Bragaglia et al. 2011). Vesperini et al. (2010), Bekki (2011), and  Conroy \&
Spergel (2011) make clear predictions on the presence/absence of  MPs and the
kind of abundance variations in globular clusters based on their mass and 
formation environment. 

Spreads in iron abundances [Fe/H] have been found so far only in the most
massive Milky Way globular clusters: in $\omega$ Cen ([Fe/H] range more than 1.5
dex, with discrete components: e.g., Norris \& Da Costa 1995; Lee et al. 1999;
Pancino et al. 2000; Hilker \& Richtler 2000; Sollima et al. 2005; Stanford et
al. 2006, 2007; Villanova et al. 2007; Johnson \& Pilachowski 2010; Marino et
al. 2011a); in M~54 (spread in metallicity of about 0.2 dex: suspected from the
colour-magnitude diagram by Sarajedini \& Layden 1995, confirmed with Ca triplet
measures by Bellazzini et al. 2008 and high-resolution spectroscopy by Carretta
et al. 2010b,c); in M~22 (spread of about 0.15 dex in metallicity, derived from 
high-resolution or Ca triplet measures: Marino et al. 2009, 2011b; Da Costa et
al. 2009); in Terzan 5 (two components, separated by about 0.3 dex in Fe:
Ferraro et al. 2009; however, Origlia et al. 2011 propose that it is not a true
globular cluster); in NGC~1851 (spread in Fe of about 0.08 dex: Yong \& Grundahl
2008; Carretta et al. 2010d, 2011; see however Villanova et al. 2010); and in
NGC~2419 (spread of about 0.2 dex in Ca: Cohen et al. 2011). This is another
clue for the formation mechanism(s).

In a few globular clusters, also internal dispersion in heavy (neutron-capture)
elements has  been found, but the exact connection with multiple populations is
not understood yet. Recent examples are NGC~1851 (Yong \& Grundahl 2008;
Carretta et al. 2011a), where $n$-capture elements variations seem to correlate
with the light elements involved in the typical globular cluster
(anti)correlations (such as Al, Na) and where also a variable C+N+O abundance is
found (Yong et al. 2009), or not found (Villanova et al. 2010);  and M22 (Marino
et al. 2009, 2011b), where $n$-capture elements correlate with Ca and Fe.

When we talk about photometric evidence of multiple populations, apart from
$\omega$~Cen, with its spectacular multiple RGBs visible in all visual
colour-magnitude diagrams (e.g., Lee 1999; Pancino et al. 2000) or M~54
projected on the Sgr dSph population (e.g., Siegel et al. 2007), generally only
subtle effects are expected on the RGB, difficult to disentangle from errors or
differential reddening. That is, unless the ultraviolet bands are involved, or
even better if we consider the Str\"omgren filters. Sbordone et al. (2011) used
theoretical computations to show the effects  of different chemical compositions
in the observed colour-magnitude diagrams in the Johnson and  Str\"omgren
systems. This is naturally explained by the influence of different (bimodal) N
abundance, since NH and CN bands are present in the ultraviolet and blue
filters. Broadening or splits have been observed in several cases when the
Johnson U filter was available, see e.g. NGC~1851 (Han et al. 2009), NGC~3201
(Kravtsov et al. 2010), NGC~288 (Rho et al. 2011), but  curiously not in IC~4499
(Walker et al. 2011). The effect is even stronger in the intermediate-band
Str\"omgren photometry, which has long been used to derive C and N abundances
(see e.g., Anthony-Twarog et al. 1995 for M~22; Hilker \& Richtler 2000 for
$\omega$ Cen). Spreads, separations, secondary sequences have been found for
several clusters. Lee et al. (2009), using a narrow-band filter centred on the
Ca {\sc ii} H+K lines, claimed not only that all the eight globular clusters
examined showed a clear indication of multiple populations (with the exception
of NGC~6397), but that there was evidence of enhanced Ca, i.e., of pollution due
to supernova explosions. This is controversial (e.g., Carretta et al. 2010a),
except for M~22 and NGC~1851. We do not have a complete explanation for these
sequences; for instance, in M~22 (Marino et al. 2011b) and NGC~1851 (Villanova
et al 2010; Carretta et al 2011a), a secondary RGB sequence seems to be 
occupied only by stars rich in $n$-capture elements. On the other hand, the 
connection between N (i.e., colours involving the bluer bands) and Na abundance
has been proven by several analyses. Marino et al. (2008) showed that Na-rich
(N-rich) and Na-poor stars in M~4 do separate in colour on the RGB if the
Johnson U filter is used. The same was found for NGC~6752 and other clusters
(e.g., Carretta et al. 2009a, 2011b; Milone et al. 2010; Lardo et al. 2011;
Marino et al. 2011b). It has also been recognised that there is a radial
dependence of the populations: stars that can be assigned to the second
generation on the basis of various properties (abundances, colours, etc.) seem
to be more centrally concentrated, see e.g. Bellini et al. (2009: $\omega$ Cen),
Kravtsov et al. (2010, 2011: NGC~3201, NGC~6752); Lardo et al. (2011: nine
globular clusters from the Sloan all-sky survey); this is in good agreement with
what is expected from the formation models (see Sect.~\ref{scenarios}). 

Finally, a word about He. He enrichment is expected to characterize the matter
exposed to hot-temperature $p-$capture processes, so that second generation
stars (i.e., O-poor, Na-rich) are supposed to be He-enhanced, compared to first
generation (i.e., O-rich, Na-poor) ones. This is indirectly confirmed by various
methods. We already spoke of the multiple MSs (see Sect. \ref{ms}). Bragaglia et
al. (2010a) found that the luminosity of the RGB bump, which should increase
with He abundance, is fainter in first generation than second generation in 14
globular clusters. Nataf et al. (2011) found that the RGB bump is brighter at
the centre of 47~Tuc, where second generation stars should dominate. The colours
of second generation RGB stars are bluer than those of first generation ones, in
agreement with expectations for a higher He content (e.g., Bragaglia et al.
2010a; Sbordone et al. 2011). Finally, He-rich HB stars are expected to be
bluer: the colours of HB stars have been examined by many authors (see e.g.
D'Antona et al. 2005; Lee et al. 2005; Gratton et al. 2010, see also next
Section). However, the direct measure of He content in RGB stars is a very
recent achievement. Dupree et al. (2011) measured equivalent widths of the He
{\sc i} line at 10830~\AA\ in 12 RGB stars in $\omega$~Cen, finding a positive
correlation with Na, Al abundance. Pasquini et al. (2011) employed a more
quantitative approach and, using also models for the atmosphere and
chromosphere, compared the strength of the same He line in two giants in
NGC~2808, one Na-rich, one Na-poor, finding the spectra compatible with a
$\Delta{\rm Y}\ge 0.17$, in good agreement with what is expected in this
cluster. 

\subsection{Horizontal branch \label{hb}} 

Evidence of multiple populations along the HB of a few globular clusters was
found already several decades ago (e.g., Harris 1974; Walker 1992; Sosin et al.
1997; Rich et al. 1997; Ferraro et al. 1998; see however the warning by Catelan
et al. 1998 about possible spurious detections). This splitting of the HB has
long been related to the second parameter problem (that is, the colour of the HB
is determined not only by metallicity, Sandage \& Wildey 1967; van den Bergh
1967), but it was only quite recently that a plausible cause was identified in
variations of the He content related to the multiple populations (D'Antona et
al. 2002, D'Antona \& Caloi 2004) following much earlier suggestions (Sandage \&
Wildey 1967; Norris et al. 1981). Variations in the abundances of the
proton-capture process elements among different stellar generations in globular
clusters are expected to be correlated with variations of the He content,
themselves invoked to explain the multiple MSs seen in mono-metallic globular
clusters such as NGC~2808 (Piotto et al. 2007; see Sect.~\ref{ms}). Since
He-rich stars burn H at a faster rate than the He-poor ones while on the MS,
they run out of fuel earlier, so their progeny on the HB are expected to be less
massive and bluer than the progeny of He-normal stars if they lose similar
amounts of mass on the RGB (D'Antona et al. 2002; see also Norris et al. 1981
for a similar argument). This has been studied in detail for individual clusters
by Caloi \& D'Antona (2005, 2007, 2008), D'Antona \& Caloi (2008); D'Antona et
al. (2010a, 2010b); and di Criscienzo et al. (2010b). Carretta et al. (2007d)
and Gratton et al. (2010a) discussed the general correlation between the
abundances of $p-$capture elements in RGB stars and the colour of HB stars, and
found indeed that a close relation exists between the extension of the Na-O and
Mg-Al anticorrelations, and that of the HB, precisely as expected in this
scenario. These studies also showed that the main parameter driving the whole
phenomenon of multiple populations is the cluster mass, confirming similar
previous results (Recio-Blanco et al. 2006; Carretta et al. 2006, 2009a) and
that the second parameter can be explained almost entirely by a combination of
the ages and masses of globular clusters, although other authors prefer age and
central density (e.g. Fusi Pecci et al. 1993; Dotter et al. 2010), itself
loosely correlated with mass.

While this is extremely intriguing, this scenario should be confirmed by the
direct determination of the chemical composition of HB stars. The surface
composition  of the hottest HB stars, with temperatures $>11,500$~K (the
location of a  distinct gap along the HB, the so called $u-$jump: Grundahl et
al. 1999; Caloi 1999), are known to be heavily influenced by sedimentation
caused by diffusion and by radiative pressure effects (Michaud et al. 1983) as
indicated by observations (e.g., Greenstein \& Sargent 1974; Crocker et al.
1986; Peterson et al. 1995; Behr et al. 1999, 2003; Fabbian et al. 2005; Pace et
al. 2006). Results are then difficult to interpret. Exceptions are the hot stars
with $T_{\rm eff}>30,000$~K (Moehler et al. 2007, 2011), which are likely a mix
of the progeny of the most He-rich stars and of late hot-flashers (that is,
stars  which lose so much mass that they leave the RGB before the core He-flash.
They  could themselves be a progeny of very He-rich stars or the result of the
evolution in binary systems). However, abundances for stars with $T_{\rm
eff}<11,500$~K can be reliably determined from spectra, although the low S/N 
ratios of the spectra available in the past prevented sufficiently accurate
determination of the abundances of the elements of interest here (He, O, Na).
Peterson et al. (1995) first obtained O abundances for blue HB stars in M~3,
M~13, and NGC~288 which decrease with increasing stellar temperature in each
cluster, in general agreement with the multiple population scenario. Recently, a
few investigations tried to determine the abundances of O and Na, and even of
He, in cool BHB stars. The restricted temperature range 9,000-11,500~K is the
only ones where reliable He abundances can be determined along the HB. For He,
the S/N required to test the multiple population scenario is very high so that
this analysis is possible only for a few close globular clusters (for instance,
the low S/N of the spectra analysed by Behr et al. 2003, have errors that are
too large), while a lower S/N can be accepted for Na and O.  These limitations
are important, however the pioneering study of seven cool BHB  stars in NGC~6752
by Villanova et al. (2009) showed that the five stars where He  could be
measured have normal He content and are Na-poor and O-rich, as expected.  Two
other studies (Marino et al. 2011c; Villanova et al. 2011) focused on M~4, the 
nearest globular cluster. This cluster has a much redder HB than NGC~6752, so
that stars on both  the red and blue sides of the RR Lyrae instability strip
could be studied. Again, as expected, the red HB stars were found to be Na-poor,
O-rich, while the blue ones are Na-rich, O-poor. 

Finally, Gratton et al. (2011b) obtained spectra for several tens of stars on
the HB of NGC~2808, which is the prototype mono-metallic, multiple stellar
population cluster. While the S/N of the spectra is not sufficiently high to
measure He abundances, Na and O abundances could be determined for several RHB
stars and six BHB stars. As expected for the multiple population scenario
described in the first paragraph of this section, the  BHB stars are O-poor,
Na-rich, and the RHB ones are O-rich, Na-poor. This strongly supports the
identification of the BHB as the progeny of the intermediate MS, and of the RHB
with the rMS, made by D'Antona et al. (2005). No reliable abundances could be
obtained for the extremely blue HB stars, that should be the progeny of the bMS
stars.

\subsection{Asymptotic Giant Branch (AGB)\label{agb}} 

Stars with strong CN-bands are very common along the RGB of globular clusters.
It is then quite curious that most AGB stars in globular clusters have weak
CN-bands, a fact first discovered by Norris et al. (1981) in NGC~6752 (see also
the review by Sneden et al. 2000). Similar results were obtained for a large
sample of giants in  M~13 by Pilachowski et al. (1996): stars with high [Na/Fe],
dominating the upper RGB, do not seem to be represented among the AGB sample.
Campbell et al. (2010) reported initial results of a medium-resolution
spectroscopic survey of about 250 AGB stars over nine globular clusters,
confirming the lack of CN-strong AGB stars in NGC~6752. In other globular
clusters (such as M~5: Smith \& Norris 1993; and possibly 47~Tuc: Mallia 1978;
Campbell et al. 2006) a few CN-strong AGB stars are present, although they are
less frequent than among RGB stars.

This low incidence is striking, since the presence of strong CN bands is usually
assumed to be evidence of more mixed material. Norris et al. (1981) proposed
that this is related to the large He abundances likely associated with the
CN-strong stars. Given their lower mass, a significant fraction - in some
clusters perhaps all - CN-strong stars do not ascend the giant branch for a
second time. This possible explanation is supported by models of HB stars (e.g.,
Sweigart \& Gross 1976; Greggio \& Renzini 1990; Lee et al. 1994).

As discussed in Sect.~\ref{hb}, the extension of the HB to low masses observed 
in several globular clusters is interpreted as the effect of enhanced He
abundances (which combine with differences in ages and possibly other parameters
to explain the so-called second parameter problem), exactly as proposed by
Norris et al. (1981) although with a different scenario for the formation of
globular clusters (see D'Ercole et al. 2008; Carretta et al. 2010a; Bekki
2011). 

Within this framework, we should then expect that there is a close relation 
between the extension of the HB (as represented e.g., by the minimum mass along 
the HB: see Gratton et al. 2010a) and the frequency of AGB stars (as given by 
the ratio between the number of stars along the AGB and those on the RGB: see 
discussion in Sneden et al. 2000). M~13 seems indeed to have a low frequency of 
AGB stars (Caputo et al. 1978; Buzzoni et al. 1983). A more comprehensive
analysis has been presented by Gratton et al. (2010b), who derived the relative
frequency of AGB stars in 26 globular clusters with extensive and uniform high
quality data and small reddening. They found a good correlation between the
ratio of AGB to RGB stars and the minimum mass of stars along the HB, with a
further dependence on metallicity. This agrees with the expectation that the
less massive HB stars ($M<0.54~M_\odot$) do not even begin their AGB phase. For
the remaining HB stars, the length of the AGB lifetime is a function of their
mass and metallicity. 

Since the mass of the HB stars is expected to be correlated with their chemical 
composition (see Gratton et al. 2010a), we then expect that most He-rich
(Na-rich, O-poor) stars in globular clusters do not reach the AGB. He-poor
(Na-poor, O-rich) stars may have an extended evolution on the AGB, up to the
luminosity of the tip of the RGB. The stars having an intermediate composition
begin their AGB phase, but terminate their evolution before reaching these
bright luminosities.

\section{Nucleosynthesis and multiple populations \label{sec:3}} 

\subsection{Observed nucleosynthesis in globular cluster environment \label{nucleosynthesis}}

We showed in the previous section that the chemical composition of globular
cluster stars  reveals a peculiar nucleosynthesis that is confined only to the
globular cluster environment. While the abundances of heavy nuclei ($\alpha$-,
iron-peak elements) are fairly constant in {\it almost} any well-studied
globular cluster (e.g., Gratton et al. 2004), the abundances of light elements
(from Li to Si) show large star-to-star variations in {\it almost} all studied
globular clusters (Gratton et al. 2004; Carretta et al. 2009a,b, and references
therein). These alterations are not shared by field counterparts with similar
metallicities (Gratton et al. 2000). 

Traditionally, elements formed via $n-$capture were thought to show neither
large star-to-star variations (Armosky et al. 1994; James et al. 2004; Yong et
al. 2006, 2008a) nor correlations with the abundance variations of light
elements. Recently, $r-$process abundance variations have been found within
metal-poor globular clusters (Roederer 2011; Roederer \& Sneden 2011; but see
Cohen 2011), although they are not correlated with the spreads in light
elements. D'Orazi et al. (2010a) showed that Ba (mostly an $s-$process element)
is usually characterised by a single abundance value in many (usually more
metal-rich) globular clusters. Since this effect seems unrelated to the spreads
in light elements, we are tempted to attribute it to original inhomogeneities in
the interstellar matter which formed the globular clusters. This might be a
counterpart of the large spread in $r-$process elements found among extremely
metal-poor stars (McWilliam 1995). Only in a few clusters, interestingly those
where an [Fe/H] spread is found, variations in $s-$process elements, positively
correlated with the Fe abundance are also detected ($\omega$~Cen: Johnson \&
Pilachowski 2010, D'Orazi et al. 2011 and references therein; M22: Marino et al.
2011b; NGC~1851: Carretta et al. 2011). In the peculiar cluster NGC~1851,
correlations between the abundance variations of light and $s-$process elements
are also observed (Yong \& Grundahl 2008; Villanova et al. 2010; Carretta et al.
2011a; see Sect.~\ref{rgb}).

As shown in Sect.~\ref{sec:2}, the most striking signature detected in almost
every globular cluster is the well known Na-O anticorrelation. This feature is
so widespread among globular clusters that must be necessarily related to their
very same origin, as suggested by Carretta (2006). With the recent survey with
FLAMES@VLT conducted by our group (e.g., Carretta et al. 2006, 2009a,b,c) it was
possible to gain an order of magnitude in statistics and to {\it quantitatively}
assess for the first time that only about one third of the currently observed
stars show primordial composition, whereas most of the stars show the modified
and quite peculiar chemistry discussed in the previous sections. Carretta et al.
(2007, 2010a) quantified the extension of the Na-O anticorrelation, showing that
it is a function of the total absolute magnitude $M_V$\ and metallicity [Fe/H]
of the globular clusters; furthermore, it also correlates with the helium
abundance (Gratton et al. 2010a). Carretta et al. (2010b) were the first to
demonstrate that the Na-O  anticorrelation is more extended at intermediate than
at low metallicities in  $\omega$ Cen and M~54, the two most massive globular
clusters in the Milky Way, both showing a large  dispersion in iron abundance.
This was later confirmed by the extensive datasets of Johnson \& Pilachowsky
(2010) and Marino et al. (2011a) for $\omega$~Cen.

The peculiar chemistry observed in globular clusters bears the clues to its
origin. Already early observations in unevolved globular cluster stars of large
star-to-star variations in N, anticorrelated with smaller spreads in C, showed
that the sum C+N increases as C decreases (e.g., Cohen et al. 2002; Briley et
al. 2004a,b; Briley \& Cohen 2001). This suggests in turn that we are not simply
seeing a conversion of C into N, and either O must be involved in the process or
variable amount of N must be considered. However, when the complete set of C, N,
O abundances is available (not a trivial experimental task) in globular cluster
dwarfs (Carretta et al. 2005) or giants (e.g., Ivans et al. 1999; Villanova et
al. 2010), the C+N+O is found to be approximatively constant (but see the
results by Yong et al. 2009 for NGC~1851). The conclusion is that a hot
H-burning environment is required to alter the abundance of light elements with
the simultaneous action of $p$-capture reactions of the CNO, NeNa, and MgAl
chains (Denisenkov \& Denisenkova 1989; Langer et al. 1993). The high
temperature required for this processing to occur in stellar interiors ($\ge
20\times 10^6$ and $\ge 70\times 10^6$ K for the NeNa and MgAl reactions,
respectively) necessarily implies a multi-generations scenario for the cluster
formation. In fact, the present-day low-mass stars cannot reach these high
temperatures; therefore the observed composition changes {\it must} be caused by
a previous generation of more-massive stars which polluted the intra-cluster
medium with nuclearly processed matter (e.g., Gratton et al. 2001), but in most
cases not with the product of supernovae.

The alterations cannot be limited to surface contamination (Cohen et al. 2002),
or they would be diluted after the first dredge-up. Instead, the C-N
anticorrelations are observed all the way from the MS to the RGB tip. Also the 
Na-O and Mg-Al anticorrelations do not significantly change with evolutionary 
status of the stars, see e.g., the case of 47~Tuc where Na and O were measured
for large samples of RGB and MS stars by Carretta et al. (2009a) and  D'Orazi et
al. (2010b), respectively, or NGC~2808 where Bragaglia et al. (2010a, see
Sect.~\ref{ms} and Fig.~\ref{f:n2808}) showed that the two unevolved stars on
the bMS and rMS seem to share the same extension of the Mg-Al anticorrelation of
giants (Carretta et al. 2009b). Thus, the overall chemical composition of
globular cluster stars demonstrates that most of the stellar structure is
affected. Moreover, we know that a large fraction of each globular cluster is
affected by this peculiar nucleosynthesis, because of the observed ratio of
polluted $vs$ primordial stars, the latter forming the living remains of the
generation which provided the processed matter. In turn, this poses severe
constraints on the mass budget of the polluters (see below).

\subsection{The nature of the polluters \label{polluters}}

The observed nucleosynthesis in globular clusters drives the selection of
suitable polluters. The best candidates must be able to modify only the light
elements without producing heavy elements such as iron (the observed intrinsic
spread in [Fe/H] is generally $<0.05$~dex, e.g., Carretta et al. 2009c). The
most popular kind of  polluters are intermediate-mass AGB (Cottrell \& da Costa
1981; D'Antona et al. 1983; Ventura et al. 2001) and/or
super-AGB\footnote{Super-AGB stars are defined as objects with initial masses in
the range 9-11~$M_\odot$ that undergo off-centre carbon ignition in partially
degenerate conditions, and end up their evolution as O-Ne white dwarfs.} stars
(Pumo et al. 2008; Ventura \& D'Antona 2010, 2011), and fast rotating massive
stars (FRMS, Norris 2004; Maeder \& Meynet 2006; Prantzos \& Charbonnel 2006;
Decressin et al. 2007a,b). For each class the model allows for (i) a hot
H-burning environment (hot bottom burning at the base of the convective envelope
for AGBs and in the central core for FRMS), (ii) a mechanism to bring the
nuclearly processed material to the surface (convection for AGBs, rotational
mixing for FMRS), and (iii) a way to release this material to the intra-cluster
medium at sufficiently low velocity to avoid escaping from the cluster potential
well (slow AGB winds and envelope ejection or slow mechanical equatorial wind
for FRMS). Alternative models have been proposed, but they seem to attract less
support. Massive stars in close binaries (de Mink et al. 2009) ejecting enriched
material at low velocity in the non-conservative evolution of the system were
criticised by D'Antona \& Ventura (2010) because this scenario cannot account
for the high fraction of second generation stars observed in globular clusters.
If the models by Decressin et al. (2009) correctly describe the impact of
rotation on the nucleosynthesis of AGB models, the AGB scenario should be
discarded, as these rotating models predict a large increase in the total C+N+O
sum  (due to the production of primary N) clearly contradicted by current
observations.

Most of the active debate is centred on the two above mentioned main candidate 
polluters, FRMS (20-120~$M_\odot$) and intermediate mass AGB stars
(4-11~$M_\odot$); it is even  possible that both mechanisms are active, though
perhaps only in a minority of globular clusters, as suggested by Valcarce \&
Catelan (2011). In an elegant paper, Prantzos  et al. (2007) used simple
nucleosynthesis calculations to define a narrow  temperature range for
H-processing (70-80~MK) where the $p-$capture reactions  could simultaneously
match the whole observed abundance pattern, from C to Mg, in the well studied
cluster NGC~6752. According to the required temperature (a strong constraint,
since it depends only on the nuclear rate), both massive AGB and the most
massive MS stars can reach the high values needed to make them suitable
candidate polluters of the star-forming gas in globular clusters. We however
note that current models for both AGB and FRMS are not able to reproduce the
exact pattern of abundances observed (see Carretta et al. 2009b), and some fine 
tuning of the most relevant parameters (nuclear cross sections, efficiency of 
convection, etc.) is required (see e.g., Ventura et al. 2011 and Karakas 2010). 
For instance, AGB stars have difficulties reproducing the severe O-depletion 
observed in the most extreme globular cluster stars ([O/Fe]$<-0.5$: Ventura \&
D'Antona 2009), while only the most massive FRMS ($M>60~$M$_\odot$) are able to
reach these low values (Decressin et al. 2007). 

Concerning the nucleosynthesis, the obvious bottom line is that polluted stars
must be He-enriched, since He is the main outcome of H-burning. There are many
indications for He abundance variations: multiple MSs (Sect.~\ref{ms}), extended
HBs (Sect.~\ref{hb}); changes in the mean abundances of Na and O observed at the
RGB bump magnitude (Carretta et al. 2007b; Bragaglia et al. 2010a,
Sect.~\ref{rgb}),  interpreted as the result of different bump luminosities for
He-normal and He-rich stars (Salaris et al. 2006); direct measures of He
abundances in RGB (Sect.~\ref{rgb}) and HB stars (Sect.~\ref{hb}). Whatever the
polluters producing Na-enriched stars are, they also provide He-rich matter.

The He production occurs in the MS for both kinds of polluters. However, while
the $p-$capture elements are also produced in the same phase in the FRMS, their 
production is decoupled from that of He for AGB stars. In $\sim 3-8~M_\odot$\
stars, He is brought to the envelope by the second dredge-up (Becker \& Iben
1979, see also Ventura et al. 2002 and Renzini 2008), while the light elements
are produced in $p-$capture reactions in hot bottom burning (Scalo et al. 1975;
Bl\"ocker \& Sch\"onberner 1991; Sackmann \& Boothroyd 1992) during the AGB
phase. While FRMS are able to produce the extremely large He abundances
($Y>0.35$) probably required to explain the blue MS of $\omega$~Cen and NGC~2808
(Bedin et al. 2004; Norris 2004; Piotto et al. 2005; D'Antona et al. 2005;
Piotto et al. 2007),  it is difficult to produce $Y>0.35$\ with AGB stars,
although values up to $Y\sim 0.38$\ can be possibly produced by super-AGB stars
(Siess 2007; Ventura \& D'Antona 2011). 

In both cases we however expect that He is produced as a primary element, and
that its production is fairly independent on metallicity; the amount of He that
can be returned to the interstellar matter is a function of the stellar mass. On
the other hand, the temperature (efficiency) of the H-burning where light
elements are forged depends on metallicity, being higher in lower metallicity
stars. Therefore, the same Na/O ratio is given by smaller mass polluters in
lower metallicity globular clusters. Observationally, if we consider only stars
with similar Na/O abundance ratios (for instance, the intermediate I component
of Carretta et al. 2009a), we should then expect a correlation between the He
production (e.g., the difference in colour on the RGB) and metallicity. This
agrees with the recent study by Bragaglia et al. (2010b). However, there is
considerable uncertainty in this result and more precise data are required to
confirm it.

More insight concerning the properties of the polluters can be obtained coupling
the NeNa and the MgAl cycles. The benchmark sample is presently the Carretta et
al. (2009b) one, with homogeneous abundances of O, Na, Mg, Al, and Si for more
than 200 RGB stars in 19 globular clusters, based on high-resolution UVES
spectra. The Na-Al correlation varies from cluster-to-cluster. In some cases
(e.g., M~4, Marino et al. 2008; Carretta et al. 2009b) Al is almost constant
over about 0.6 dex of change in Na, whereas in other cases (e.g., NGC~2808)
large variations in Al are also observed. The maximum Na abundances present
instead only a small spread, among the 19 clusters. This indicates that only a
subset of polluters responsible for the Na, O variations was able to reach
temperatures sufficiently high to activate the MgAl cycle. The ``leakage" from
the MgAl cycle on Si (Yong et al. 2005), producing a Mg-Si anticorrelation and a
Si-Al correlation, observed in several clusters (Carretta et al. 2009b), points
to temperatures in excess of $65\times10^6$~K (see Arnould et al. 1999; Ventura
et al. 2011).

The production of Al was found to depend on the [Fe/H] and $M_V$\ of globular
clusters, similarly to the extreme O processing (Carretta et al. 2009a),
indicating that the polluters properties change regularly with cluster
luminosity and metallicity. Unfortunately, only a qualitative agreement seems
reached by current theoretical predictions of nucleosynthesis, and only after
some adjustments of reaction rates of key ingredients (e.g., the
$^{22}$Ne(p,$\gamma$) for Na in AGB models, Ventura \& D'Antona 2008; and by a
large factor the $^{24}$Mg(p,$\gamma$) rate for FRMS, Decressin et al. 2007a). A
promising diagnostic seems to be the output from the MgAl cycle. If the scenario
of FRMS is valid, we should expect more than a single gas pool (as predicted by
the AGB scenario, see D'Ercole et al. 2008, and below in this Section), because
in the original FRMS scenario, second generation stars were supposed to form
around each polluter, with a range of yields depending on the mass and rotation
rate. In this case, a broader Na-Al distribution, instead of a uniparametric
relation, should be expected. This is what is possibly seen for a few globular
clusters (e.g., NGC~3201, NGC~6254, Carretta et al. 2009b; M~3, Sneden et al.
2004), while for other globular clusters a single relation is reasonable, so
more extensive studies are required. Note, however, that in the latest version
of the FRMS scenario (Decressin et al. 2010) the idea of second generation stars
formed around each individual star is abandoned in favour of a gathering of all
circumstellar disks in a common reservoir at the cluster centre, similar to the
AGB scenario. 

A sensitive test for the nature of the polluter is provided by Li abundances,
since Li is destroyed at temperature of about $2.5\times10^6$~K, i.e, it cannot
survive at the temperature of hot H-burning. Therefore the ejecta of FRMS must
be Li-free. On the other hand, some Li production is possible for AGB stars
through the Cameron \& Fowler (1971) mechanism, so that the lack of Li-O
correlation or a slope clearly different from unity would only be consistent
with the AGBs. The Li-O correlation found by Shen et al. (2010) in NGC~6752 is
statistically significant, but they observed less Li depletion than O depletion.
A similar result has been obtained for M~4 (Mucciarelli et al. 2010; D'Orazi \&
Marino 2010; Monaco et al. 2011). In these globular clusters the polluting
matter seems Li-enriched, favouring the AGB scenario. It is however curious that
the Li yields for AGB stars are similar to the Spite plateau level (D'Antona \&
Ventura 2010); these yields are sensitive to the adopted mass loss law.

Finally, the study of heavy elements from $s-$process nucleosynthesis may
provide important clues regarding the nature of first-generation polluters. FRMS
may produce at most only the lightest of these elements (see however the claim
by Chiappini et al. 2011). The {\it main} component is instead synthesised in
the AGB phase of low mass stars ($\sim 1.5-3 M_\odot$). Despite the widespread
constancy of $s-$process elements observed in most globular clusters, there are
exceptions. Carretta et al. (2011) confirmed and extended the findings by Yong
\& Grundahl (2008) of a strict correlation between $s-$process dominated
elements and light elements involved in $p-$capture reactions in NGC~1851. A
similar result has been obtained for M~22 by Marino et al. (2011b). Since in
these globular clusters an overabundance in the total C+N+O is observed for the
same stars (but see Villanova et al. 2010 for the constancy of this sum), these
observations have been interpreted as convincing evidence of pollution by AGBs.
However, there might be a significant timescale problem, since only the most
massive AGBs (4-8~$M_\odot$) and the super-AGBs (Pumo et al. 2008; Ventura et
al. 2009) contribute to the enrichment/depletion in light elements, on a time
range between 40 and 160 Myr (Schaller et al. 1992). On the other hand, the
timescales involved in low-mass producers of $s-$process elements are of 340 Myr
up to almost 2~Gyr. This implies a time delay of several $10^8$~yr to be added -
rather {\it ad hoc} - in the phase of pollution from AGBs, which should  instead
run smoothly. In these cases, the abundance of F, an element involved in 
$p-$capture reaction, but only poorly studied in globular cluster (Smith et al.
2005; Yong et  al. 2008b) may offer precious insight: from theoretical
predictions, F is formed in the He intershell of low mass AGB, that are also
responsible for C+N+O and $s-$process variation, while it is destroyed by
H-burning at high temperature. Measuring F abundances in RGB stars belonging to
M~22 and NGC~1851 could then reveal the precise mass range of the polluters at
work in these globular clusters. 

\subsection{Main scenarios of globular cluster formation \label{scenarios}}

The main candidate polluters are currently used to design scenarios and models
for the globular cluster formation. D'Ercole et al. (2008) made one of the first
set of N-body and hydrodynamical simulations to show that the ejecta of the most
massive (4.5-8~$M_\odot$) AGB stars of the first generation can collect in a
cooling flow at the centre of the cluster, where a second generation forms.
D'Ercole et al. (2010) explored in detail the constraints imposed by
spectroscopic observations on a large number of tunable parameters, to show how
the interplay of many ingredients may reproduce the observed chemical pattern of
globular clusters. Other models whose main players are AGB stars were proposed
by Bekki et al. (2007), Conroy \& Spergel (2010), and Bekki (2011). A
qualitative scenario to explain the relation of globular cluster chemistry and
global parameter was sketched by Carretta et al. (2010a), where all current
globular clusters formed within a more massive {\it precursor} that was
afterward shredded by strong interactions with the main galaxy, and by Valcarce
\& Catelan (2011) who used a toy model to qualitatively explain the observed
colour-magnitude diagrams of globular clusters, assuming that the main parameter
at work is the initial mass of the precursor. On the other hand, the FRMS were
used at first to depict simple scenarios for the origin of globular clusters
(Decressin et al. 2007a,b) and recently included in more sophisticated N-body
models, taking into account the dynamical effects of early gas expulsion
(Decressin et al. 2010). Many pros  and cons are discussed in depth in the
original and in summary papers (such as  Renzini 2008, Martell 2011); here we
examine only the main open problems faced by these models.

Common to all scenarios is the ``mass budget problem". The observations (e.g.,
Smith \& Norris 1982; Carretta et al. 2009a) show that the residual first
generation component observed in globular cluster is much less than the second
generation formed by ejecta of its massive stars. If a canonical IMF is adopted,
the total mass of AGB ejecta in a globular cluster is only 1-10\% of the total
globular cluster mass (Bekki \& Norris 2006); similarly, if slow wind from FRMS
stars is used to accumulate polluted mass, this provides only about 10\% of the
total number of low-mass stars (Decressin et al. 2007b).

A first way out could be to assume a non canonical, top-heavy IMF for the first
generation (D'Antona \& Caloi 2004; Bekki \& Norris 2006; Decressin et al.
2007a,b). The second solution was proposed for the first time by Bekki et al.
(2007): the globular clusters must have been initially a large factor (10-100)
more massive than today and almost all the first generation stars must have been
preferentially lost. This may occur as response to early mass loss (e.g.,
D'Ercole et al. 2008; Vesperini et al. 2010;  Decressin et al. 2010) for which
gas expulsion by supernovae is an ideal  candidate. The induced change in the
potential well unbinds stars in the outer  part of the cluster (mostly first
generation stars, because the second generation is more centrally concentrated,
especially such early in a cluster lifetime), generating the preferential loss.
This approach is currently adopted by all groups (D'Ercole et al. 2008, 2010;
Decressin et al. 2007, 2008, 2010; Carretta et al. 2010, Gratton \& Carretta
2010; Conroy 2011; Schaerer \& Charbonnel 2011; Bekki 2011). Were the initial
mass of a globular cluster similar to the one currently observed (Conroy \&
Spergel 2010), the degree of pollution would be insufficient to reproduce the
observations (D'Ercole et al. 2011). Alternatively, the globular clusters could
be born within larger systems, such as ancestral dwarf galaxies, that
contributed the yields of a much larger AGB population (Bekki et al. 2007) to
form the second generation (see also B\"oker 2008; Carretta et al. 2010a; and
Bekki 2011 for this particular scenario).

The two main scenarios share a few common features. In particular, both predict
that second generation stars should be initially more centrally concentrated. In
the D'Ercole et al. (2008) model this naturally follows from the concept of
cooling flow; in the scenario by Decressin et al. (2007b), massive polluters are
born in, or migrated rapidly to, the cluster centre, so that second generation
stars are only created there. However, in the following evolution with two-body
relaxation processes, first generation and second generation stars are mixed
into a quite homogeneous structure in about two relaxation times (Decressin et
al. 2008; D'Ercole et al. 2008) and after a Hubble time any radial difference
between the two components is expected to be completely erased (Decressin et al.
2010) save for the globular clusters with the longest relaxation time. While
clear evidence of different radial distribution for  different stellar
generations has been found in a few cases (e.g., $\omega$ Cen, Sollima et al.
2005, Bellini et al. 2010; NGC~1851, Zoccali et al. 2009,  Carretta et al.
2010d; NGC~3201, Carretta et al. 2010e; and several other globular clusters,
Lardo et al. 2011), the extant data are not sufficient to give firm conclusions
on most clusters (see Carretta et al. 2009a).

Another common requirement is that of dilution of polluting gas with unprocessed
material, necessary to reproduce all the observed ranges of abundances; it also
naturally accounts for the fact that some residual gas must be found in early
globular clusters, since the efficiency of star formation cannot be 100\%.
Prantzos et al. (2007) noted that the extreme abundances in NGC~6752 could be
reproduced only after mixing the polluting matter with 30\% of unprocessed
material, with different dilution factors accounting for other less extreme
variations. Dilution with pristine material may help explaining the Li
abundances, e.g. why second generation stars with extreme processing are not
devoid of Li in NGC~6752 (Pasquini et al. 2005).

Required for the FRMS scenario, the dilution is a {\it mandatory} ingredient for
the very same survival of the AGB scenario. Until recently, the dilution
argument was used only to reproduce the typical shape of the anticorrelations,
as for the massive star scenario. In some particular cases, notably NGC~1851, it
was postulated to fine tune the AGB pollution and reproduce the difference in
C+N+O required to explain the double SGB without a large He content (excluded 
by the MS and HB properties, see Ventura et al. 2009). However, D'Ercole et al.
(2011) and D'Antona et al. (2011) clearly state that the nucleosynthesis of AGB
models actually predict that Na and O {\it correlate}, at odds with all observed
chemical patterns in globular clusters. The only way to reconcile the
observation with theory is to invoke an early - but not-too-early, see below -
dilution of the AGB ejecta with pristine gas.

Several mechanisms have been proposed to gather the pristine matter required for
dilution: gas left from early star formation, Bondi accretion and gas sweeping
from the proto-cluster orbiting a gas-rich ancestral galaxy (Conroy \& Spergel
2010), self pollution by wind of first generation stars less massive than
polluters (Gratton \& Carretta 2010), and material collapsing back from a torus
left behind by a bi-conic, collimated gas sweeping generated by supernovae in
the proto-cluster (D'Ercole et al. 2008). Most of these are critically discussed
in D'Ercole et al. (2011). In particular, if the start of dilution is
simultaneous to that of pollution, the related models (Conroy \& Spergel;
Gratton \& Carretta) are discarded, because in the model by D'Ercole et al. the
only way to produce the extreme component of second generation stars is by means
of pure massive AGB ejecta. In this case, however, a correlation between Na and
O should be observed. Of course, this last requirement does not hold in the case
of FRMS, thanks to the more favourable nucleosynthesis, able to reach even very
high He enrichment values (up to Y=0.40 and more, Decressin et al. 2007b) even
in presence of some dilution.

The distinctly separated MSs observed in $\omega$~Cen and NGC~2808 put serious
constraints on dilution and polluters. The clear separation implies that within
each sequence the matter is homogeneous, both in He and in Fe. A neat separation
was deemed impossible by Renzini (2008) for FRMS forming second generation
stars  around each massive first generation star. However, these distinct
sequences are only observed  in these two massive clusters (see however
Sect.~\ref{ms}) and the latest version of the FRMS scenario predicts a gathering
of polluted gas in the generic centre of the globular cluster (Decressin et al.
2010). A study of discreteness vs continuous  distribution in other globular
clusters would help to clarify this question. We can, for instance, study the
nature of the Na-O anticorrelation to find whether first generation and second
generation are distributed smoothly or are confined into discrete clumps.
Although the samples by Carretta et al. (2009a) are quite large, the errors
associated with the abundances derived from GIRAFFE spectra are not sufficiently
small for this kind of analysis. A few sparse studies, e.g. Marino et al. (2008)
for M~4, seem to indicate a separation between the two stellar generations, but
the road is still open to many interpretations.

\subsection{At the high-mass tail of the globular cluster mass distributions \label{tails}}

Some scenarios for globular cluster formation may be quite easily modified to
reproduce features observed at the high-mass end of globular cluster mass
distribution, notably for the  two most massive known Milky Way globular
clusters, both with a cosmic scatter in Fe abundances:  $\omega$ Cen (from
Freeman \& Rodgers 1975; Butler et al. 1978, Norris \& Da  Costa 1995 to Johnson
et al. 2010 and Marino et al. 2011a), and M~54 (associated to the Sgr dSph;
Sarajedini \& Layden 1995; Bellazzini et al. 2008; Carretta et al. 2010b,f). The
observed intrinsic spread in Fe abundances calls for repeated and large bursts
of star formation in these clusters, where all the available range in stellar
masses does contribute to the nucleosynthesis. The strong resemblance with the
behaviour observed in local dwarfs (see Tolstoy et al. 2009) led some authors
(e.g., Bekki \& Freeman 2003; Bekki et al. 2007; B\"oker 2008; Carretta et al.
2010b,f) to suggest that massive globular clusters had their initial evolution
in the core of dwarf galaxies. In both these cases the Na-O anticorrelation is
present at different metallicities, and always with a larger extension in the
metal-intermediate component (see Sect.~\ref{nucleosynthesis}): the most heavily
polluted stars belong to this component. These observations can be reconciled
with a general scenario if the burst of star formation generating the metal-rich
component was delayed by as much as 10-30 Myr, so that their core collapse
supernovae inject a notable energy input, disturbing and even suppressing the
pollution from high mass stars in the (spatially nearby?) metal-poor component
(e.g., Carretta et al. 2010f).

Another possible variant of the last scenario resumes the idea (e.g., van den
Bergh 1996) that beside the large, massive globular clusters spiralling to the
centre of their own galaxy, finding only a diffuse nuclear component (M~54 and
possibly $\omega$ Cen), another channel is feasible: the merger of two
originally distinct globular clusters in the same ancestral galaxy before the
disruption within the Galaxy (Carretta et al. 2010d, 2011a). This possibility
may explain with economy of hypothesis the evidence of bimodality found by
photometry (Milone et al. 2008) and spectroscopy (Yong \& Grundahl 2008;
Villanova et al. 2010; Carretta et al. 2011a) in the peculiar cluster NGC~1851,
characterised by a small difference of Fe and a larger spread in Ba, positively
correlated with metallicity, as the enhancement in $s-$process elements is
higher in the more metal-rich component. In this population the level of
[$\alpha$/H] is also higher, suggesting an additional contribution from
core-collapse supernovae. All these characteristics are shared with the more
metal-poor cluster M~22 (for which more limited samples are available: Da Costa
et al. 2009; Marino et al. 2009, 2011b). In these two globular clusters, both
the Fe-poor/Ba-poor and Fe-rich/Ba-rich components present an Na-O
anticorrelation, as if two individual clusters were traced. This consideration,
together with other evidence of bimodalities (in the HB and SGB distribution,
Walker 1992, Milone et al. 2008; and in the RGB splitting, Lee et al. 2009, Han
et al. 2009), strongly point toward the merger scenario. Undoubtedly, other
observations are required to fully account for the complex nucleosynthesis of
these clusters. Note, however, that a merger scenario could also explain the
so-called extended (old) star clusters recently found in a variety of galactic
environments (Br\"uns et al. 2011).

\subsection{Present-day young/massive clusters \label{young}}

The Na and O abundance variations, with large spreads up to 1 dex, allow us to
resolve age differences of a few $10^8$ yrs (in the case of AGB pollution) and
as low as some $10^6$ yrs (for FRMS). However, even this powerful diagnostic
cannot provide the exact chain of events occurring early in a globular cluster
life. For example, a typical problem faced in the FRMS scenario is the very
narrow time gap between the end of the slow wind phase by most massive stars
(about 6~Myr) and the explosion of the bulk of core-collapse supernovae (about
10 Myr after the star formation burst) when compared with the expected length of
the star formation episodes (see Sect.~\ref{intro}). Scenarios of
inhomogeneities in the interstellar matter, with  preferential escape of
supernova ejecta through tunnels (Prantzos \& Charbonnel 2006)  or of a bimodal
behaviour above a critical mass (where only the outer cluster  regions
participate in the winds, Tenorio-Tagle et al. 2007) have been proposed  to
avoid the mixing between slow wind and fast supernova ejecta. However, details
are still poorly known and more insight may be reached by observing young
massive clusters as proxy of early evolutionary phases in the formation of
stellar clusters.

A summary of the properties of young massive clusters can be found in the review
by Portegies-Zwaart et al. (2004). A promising starting point could be the study
of the best observable examples of stellar systems that could become globular
clusters  in a few Gyr, such as the two young massive Galactic clusters RSGC1
(Bica et al.  2003) and RSGC2 (Stephenson 1990). Their young ages ($12\pm2$ and
$17\pm3$~Myr,  respectively, Davies et al. 2007, 2008) could provide a fine test
on the nature  of polluters: if spreads in Na, O abundances are detected, only
rotating massive stars can have enriched them. Interestingly enough, a third
massive and obscured cluster harbouring red supergiants (RSGC3) is found
surrounded by an extended, very massive (more than $10^5 M_\odot$) association
(Negueruela et al. 2011), tantalizingly suggestive of the loose association
around the early proto-cluster in the qualitative scenario depicted by Carretta
et al. (2010a).

Bimodal or multi-modal star formation seems to be the rule in a variety of
massive star forming environments. Age spreads up to $\sim 30$ Myr are found for
some clusters in galaxies beyond the Local Group (Larsen et al. 2011), as well
as in the Large Magellanic Cloud cluster NGC~1850 (age $\sim 50$~Myr) where
younger stars ($\sim 4$~Myr) are located in a small nearby cluster (Gilmozzi et
al. 1994). A bimodal age distribution (clustered at the ranges $\sim 10-16$~Myr
and $\sim 32-100$ Myr) was found for the resolved part of the massive stellar
cluster Sandage~96 in NGC~2403, with younger (brighter) stars more concentrated
than the older component (Vinko et al. 2009). The authors however attribute this
effect to possible accretion (and subsequent star formation) from the nearby
interstellar matter (Pflamm-Altenburg \& Kroupa 2008). In an interesting study,
Greissl et al. (2010) derived the integrated properties of a massive young
cluster (89/90) in the Antennae (NGC~4038/30) galaxies through its H and K-band
spectrum. They interpreted this spectrum as a superposition of two star clusters
of different ages ($<3$ and 6-18~Myr, respectively) with a total mass of $10^7
M_\odot$ and a ratio $M_{old}/M_{young} \le 0.2$. According to the authors it is
possible that the older cluster was initially born with a high mass and lost a
large fraction of its star mass due to the dispersion into the field, as
suggested for the early evolution of currently observed globular clusters. A
recent study of stellar populations in the field of the Small Magellanic Cloud
star cluster NGC~346 by De Marchi et al. (2011a) found pre-main sequence stars
in a bimodal distribution peaked at $\sim 1$ and $\sim 20$~Myr, with a
statistically significant gap in star formation. However, the two generations
have markedly different spatial distribution, supporting the conclusion by
Hennekemper et al. (2008) of a lack of obvious sequential star formation.
Analogously, De Marchi et al. (2011b) estimated a star formation extending over
a long period in the central regions of 30~Dor in the Large Magellanic Cloud,
since one third of the pre-main sequence they found have ages around 4 Myr,
whereas the other pre-main sequence stars show ages up to $\sim 30$~Myr.
Moreover, the younger pre-main sequence stars are concentrated near the centre
of the ionising cluster R136, while older pre-main sequence stars are found more
uniformly distributed. Although separate stellar generations indicate a
sequential star formation in a broad sense, the authors do not find a clear sign
of causal effects between the two. On the other hand, evidence of  triggered
star formation in compact regions, with a general progression in age  is found
for some super star clusters in the low metallicity star forming galaxy
SBS0335-052E (Thompson et al. 2009), whose characteristics would mimic an
environment similar to star formation sites in the early Universe.

The above evidence seems to conflict with both main scenarios. On one hand, the
small time interval between massive (rotating) stars pollution and supernova
explosion requires fast star formation, whereas the observations often reveal
age spreads up to 30 Myr between two stellar generations. This may be a problem
for the FRMS scenario, unless particular models of star cluster winds are
adopted (see Decressin et al. 2010). On the other hand, the age gap is often so
short that AGBs do not have sufficient time to start polluting the new
generation. Clearly more theoretical work and additional observations are
required.

\section{The globular clusters-Milky Way connection \label{sec:4}} 

Current cold dark matter cosmological models (White \& Rees 1978; Moore et al.
1999) indicate that the Milky Way stellar halo was assembled from many smaller
systems. Although this general picture is widely accepted, there still are
unclear points, for instance, the apparent discrepancy between predicted dark
matter substructures and observed Milky Way satellites, a discrepancy that has
become known as the ``missing satellite problem" (Klypin et al. 1999; Moore et
al. 1999). The solution probably resides in the feedback of star formation (see
e.g., the discussion in Wadepuhl \& Springel 2011), although other solutions
have been proposed (e.g., Macci\`o et al. 2010). It is then clearly of interest
to probe the earliest phases of the Milky Way building up by using the fossil
records left by  these past events, such as the composition and dynamics of the
oldest and most  metal-poor stellar components. We review here the work done
recently on the connection between globular clusters and the halo (see Helmi
2008 for a recent review of the halo properties, and Brodie \& Stadler 2006 for
a review of the properties of globular cluster systems in galaxies).

That the present-day clusters are (much) less massive than at the time of
formation is common wisdom, apart from the requirements dictated by the
multi-population nature of globular clusters. A large fraction of stars should
be lost in the very initial phase of globular cluster evolution through violent
relaxation following gas expulsion and mass loss from the most massive stars
(e.g., Baumgardt et al. 2008). On a much longer timescale, a significant
fraction of stars should evaporate: two-body encounters (e.g., McLaughlin \&
Fall 2008) and other mechanisms (such as disk shocking: see e.g., Aguilar et al.
1988) move a number of stars outside the tidal radius. Some fraction of the
stars should be  removed from the cluster within a relaxation time, typically
$\sim 5\times  10^8$~yr (see the compilation by Harris 1996). A substantial
fraction of the original globular cluster mass should then have been lost, more
efficiently among smaller globular clusters which have shorter relaxation times.
There is ample evidence for this loss, either from tidal tails and extra-tidal
stars around some globular clusters (e.g., Odenkirchen et al. 2001, 2003;
Grillmair \& Johnson 2006; Belokurov et al. 2006; Chun et al. 2010; Jordi \&
Grebel 2010; Sollima et al. 2011 and references therein), or from the deficiency
of small mass stars (which are preferentially lost if energy equipartition
holds) in others (H\'enon 1969; Richer et al. 1991; De Marchi et al. 2007; De
Marchi \& Pulone 2007). Since these processes may lead to cluster dissolution on
a timescale comparable to the lifetime of the globular clusters (see e.g.,
Lamers et al. 2005), current globular clusters are the survivors of a
potentially larger initial population (Fall \& Rees, 1977; Gnedin \& Ostriker
1997). These mechanisms might have transformed a power law (Fall \& Zhang 2001;
Vesperini \& Zepf 2003; Boley et al. 2009; Elmegreen 2010) or a Press-Schechter
(Ricotti 2002) initial distribution of masses, generally found for young objects
(see the review in Gieles 2009), into the currently observed  approximately
log-normal distribution with a peak at $10^{5.3}$M$_\odot$ (see  reviews by
McLaughlin 2003 and Brodie \& Strader 2006). However, it is possible that the
mass distribution was originally log-normal (Parmentier \& Gilmore 2005, 2007). 
The lack of knowledge about the properties of the original distribution of
globular clusters, of the environmental condition (see Elmegreen 2010), and the
subtleties of models of mass loss from globular clusters (see e.g., Kruijssen
2009 and references therein) prevent an accurate estimate of the  original total
mass of globular clusters (see however Baumgardt et al. 2008 and Schaerer \& 
Charbonnel 2011, for attempts to reconstruct these values from observed 
distributions). Up to a few years ago, data adequate to study the connection 
between halo and globular clusters was insufficient. Extensive, high quality
data set probing  large samples of stars both in globular clusters and the field
are however changing this situation.

Concerning field stars, very extensive surveys are now available, mainly (but
not uniquely) thanks to the Sloan Digital Sky Survey and following surveys  (see
e.g., Carollo et al. 2007; Juric et al. 2008; Martell \& Grebel 2010; Martell et
al. 2011;  see the review by Helmi 2008). On the other hand, D'Ercole et al.
(2008, 2010), Carretta et al. (2010a), and Bekki (2011) were able to place the
properties of different stellar generations in globular clusters and global
cluster parameters into a general framework for the formation of globular
clusters. The resulting scenario is one where probably {\it all} globular
clusters, and not only the most massive ones (see Carretta et al. 2010a;
Kravtsov \& Gnedin 2005; B\"oker 2008), formed at the centre of larger systems.
These were similar to the cosmological fragments that were the starting points
for the assembly of dSphs, but the precursors of currently observed globular
clusters probably started their evolution much closer to the central regions of
the Milky Way, and the interaction between them and the whole galaxy occurred
before the chemo-dynamical evolution of these objects could proceed up to the
expulsion of most of its gas content. This agrees well with numerical
simulations (Bekki \& Chiba 2001; Boley et al. 2009; Elmegreen 2010) and leads
to several main consequences: (i) globular clusters populate regions much closer
to the centre of the Galaxy than dSphs; (ii) globular clusters lost their dark
matter content, and have low mass-to-light ratios, while dSphs are dark matter
dominated objects (e.g., Darbinghausen et al. 2008); (iii) for this reason,
globular clusters are compact, tidally limited objects, while the luminous
component of dSphs is more diffuse and occupies only the central part of their
dark matter halo; (iv) since dSphs evolved substantially as isolated objects,
their evolution is dominated by their total mass, and they obey to
mass-luminosity and mass-metallicity relations (Kirby et al. 2008). This is not
true for globular clusters (see discussion in Gratton 2008). 

Although largely qualitative, this scenario is self-consistent and is able to 
reproduce many observed characteristics of globular clusters. Attempts to put
this scenario on more quantitative basis have been made by D'Ercole et al.
(2008, 2010, 2011) and Bekki (2011). These models are exploring various relevant
questions, including the formation of the cooling flow, the interaction between
the cluster and the  interstellar medium, etc. On the other hand, Vesperini et
al. (2010, 2011)  explored the following evolution of the clusters, and showed
that many properties of the globular cluster sub-populations can be reproduced
by these models (concentration, binary fraction, etc.).

A major question is whether a (even large) part of the galactic field stars
were  formed in clusters and are now a main component of the halo. Various
authors  proposed infant cluster mortality as the major source of halo stars
(e.g.,  Baumgardt et al. 2008). Although globular clusters currently represent
only $\sim 1.2$\% of  the mass of the Galactic halo (Gratton et al. 2011b, see
next section; a higher  estimate of $\sim 2$\% was given by Freeman \&
Bland-Hawthorn 2002), this  suggestion has recently received renewed interest
because it is now clear that the bulk of stars in globular clusters belongs to
the second generation (Carretta et al. 2009a,b). This population is typical of
globular clusters (Gratton et al. 2000), so that its main signature, the Na-O
anticorrelation, might be also considered the true tracer of a {\it bona fide}
globular cluster. The small fraction (about 2.5\%) of Na-rich and/or CN-rich
stars in the field halo (Carretta et al. 2010a; Martell \& Grebel 2010) may well
come from second generation stars evaporated from globular clusters. On the
other hand, a residual of the first generation is still present in globular
clusters but only as a minority component (about one third) of the current total
population. To reproduce correctly the ratio of first generation to second
generation and the observed chemical features, a large fraction of stars of the
first generation (more than  90\%) must have been lost from globular clusters
(see e.g. Decressin et al. 2008; D'Ercole et  al. 2008, 2011; and references in
Carretta et al. 2010a).

In the remaining of this section, we discuss in more detail the possible
relation between the first generation in globular clusters and the field halo
stars.

\subsection{The mass of the globular clusters compared to the mass of the halo \label{mass}}

Gratton et al. (2011b) estimated that, if reference is made to the same volume,
that is, the region of the Galaxy dominated by the Galactic halo, about 1.2\% of
the halo stars are currently in globular clusters. While this value has a
considerable uncertainty, it may be considered as a conservative estimate. As
mentioned above, the original fraction of stars in globular clusters should have
been larger because they are expected to lose a significant part of their stars
during their evolution. For reasonable initial conditions, the rates of
evaporation of first generation and second generation stars were likely
different only in very early phases, within two relaxation times (Vesperini et
al. 2010; Decressin et al. 2008, 2010). Let us assume for simplicity that they
lost stars at the same rate after second generation stars formed. If we then
adopt the value cited above of 2.5\% of second generation stars now present in
the general field, and the present ratio of first generation to second
generation stars, we deduce that the fraction of stars that belonged to globular
clusters at the end of the formation of the second generation is 3.7\% of the
current halo stars. Since currently 1.2\% of the halo mass is still in globular
clusters, at that epoch they were four times more massive than they are now and
that about 5\% of the halo was in globular clusters. A similar argument may be
found in Schaerer \& Charbonnel (2011).

However, as mentioned in Sect.~\ref{scenarios}, the mass involved in the process
of formation of globular clusters was likely significantly larger than their
mass after formation of the second generation. Several authors (Bekki et al.
2007; D'Ercole et al. 2008, 2010; Decressin et al. 2008; Vesperini et al. 2010;
Carretta et al. 2010a; Martell \& Grebel 2010; Gratton \& Carretta 2010; Conroy
2011; Schaerer \& Charbonnel 2011; Bekki 2011) provided estimates of the initial
mass of the  primordial population of globular clusters required to satisfy the
above constraint. In spite of different assumptions, they all agree on very
large ratios (from 10 up to 25) between the original first generation population
and the globular cluster stars after formation of the second generation.

The original proto-globular clusters should then have included $>50$\% of the
halo mass. This  value further rises if part of the mass lost by the potential
polluters was not  used to form second generation stars. There is then a strong
argument suggesting that the  primordial stars of the current globular cluster
were responsible for a large fraction,  possibly the majority of the current
field halo stars, although a contribution  by dSph's, mainly to the outer halo,
is certainly present (see the discussion in Helmi 2008). Furthermore, if we
consider that the total number of globular clusters per unit halo mass does not
depend strongly on the magnitude or Hubble type of the host galaxy (Bekki et al.
2008) we can conclude that the massive star-formation episodes producing
globular clusters played a very important role in the formation of stars in
galaxies, and might have substantially contributed to the reionisation of the
universe at $z\geq 6$\ (Ricotti 2002; Schaerer \& Charbonnel 2011).

\begin{figure}
\centering
\includegraphics[scale=0.45]{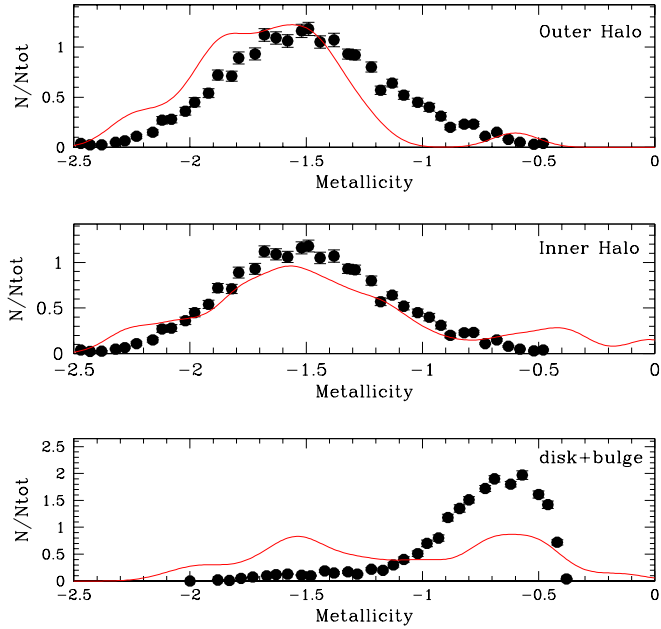}
\caption{Metallicity distribution of stars in the disk (lower panel), inner 
halo (middle panel), and outer halo (top panel) from Ivezic et al. (2008),
compared to the distribution of Galactic globular clusters. Filled circles are
field stars  from Ivezic et al., and red solid lines are the generalised
histograms for globular clusters. All distributions are normalised to unity
integrating over the whole range.}
\label{f:ivezic}.
\end{figure}

\subsection{Chemical composition \label{chemistry}}

The order-of-magnitude estimates of the previous section suggest that a  large
fraction of the Milky Way stellar halo is formed by debris of proto-clusters. 
This result is supported by the good agreement, especially for the inner halo,
between the metallicity distribution of halo globular clusters (data from the
Harris 1996  catalogue) and of the Milky Way halo, as determined from in situ
observations  (Ivezic et al. 2008; see Fig.~\ref{f:ivezic}). 

\begin{figure}
\centering
\includegraphics[bb=20 20 570 700, clip,scale=0.43]{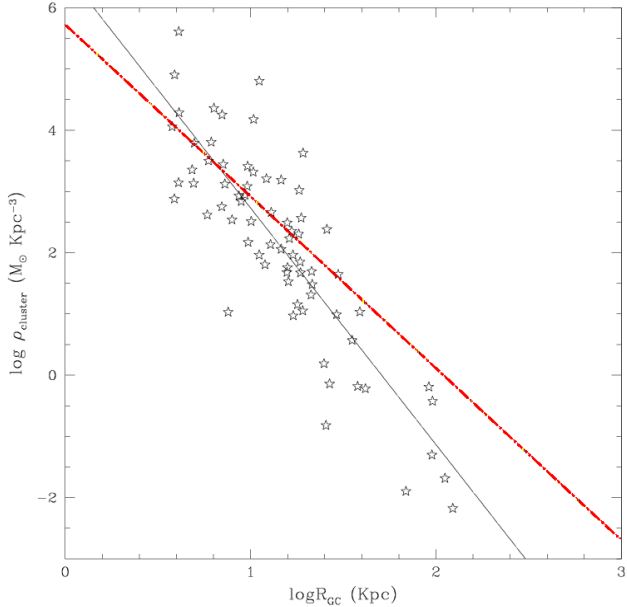}
\caption{Run of the density of stars within halo globular clusters, $\rho_{\rm
cluster}$,  with the Galactocentric distance $R_{\rm GC}$. The solid line is the
best fit of the data; the dash-dotted, red line represents the slope of the
density appropriate for the inner halo, according to Juric et al. (2008). }
\label{f:density}
\end{figure}

Element-to-element abundance ratios provide a crucial clue about the origin of 
the stars. For instance, the overabundance of $\alpha-$elements with respect to 
Fe has been used to conclude that the majority of field halo stars cannot have 
formed in objects such as present-day dSph's (see Shetrone et al. 2003; Tolstoy
et  al. 2009). On the other hand, there is no known difference between the
chemical  composition of first generation stars of globular clusters and that of
field halo stars. This is discussed  at length in the literature (e.g., Gratton
et al. 2004 and references therein;  Helmi 2008; Carretta et al. 2009a,b for O,
Na, Mg, Al, Si; Carretta et al. 2010a for Al; Carretta et al. 2010c for Ca;
D'Orazi et al. 2010a for Ba; and Lucatello et al. 2011, for Mn, Cu, and other
$n$-capture elements). Also Li abundances for first generation stars in globular
clusters and field halo ones agree well (Bonifacio et al. 2002; Pasquini et al.
2005; Lind et al. 2009; D'Orazi \& Marino 2010). On the other hand, second
generation stars clearly differ concerning the abundances of the elements
involved in high temperature H-burning, as we have seen throughout this review.

\subsection{Density distribution \label{density}}

The radial distribution of halo globular clusters matches well the distribution
of halo  stars (Brodie \& Strader 2006; Baumgardt et al. 2008; Helmi 2008). To
further  show this, we first estimated the run of the density of stars within
globular clusters with  galactocentric distance $R_{\rm GC}$. This was done by
ordering the halo globular clusters  according to increasing $R_{\rm GC}$ (see
Appendix in Carretta et al. 2010a).  This allows us to define a shell pertinent
to each globular cluster, defined by its $R_{\rm GC}$  and the same value for
the globular cluster immediately preceding. We then estimated the  density of
globular cluster stars in this shell by dividing the mass within each globular
cluster (simply  using an average $M/L_V=2$) for the volume of the shell. The
run of this density is plotted against $R_{\rm GC}$ in Fig.~\ref{f:density},
together with the best fit line and that representing the slope of the density
appropriate for the inner halo, according to Juric et al. (2008; see also Harris
1976 and  Morrison 1993, and the review by Helmi 2008 for a thorough discussion
and words  of caution about the meaning of such an average value). This second
line  represents a good fit for the inner halo, while the density slope for
globular cluster  is clearly steeper in the outer halo. These slopes agree
fairly well with the  predictions given by numerical simulations by Steinmetz \&
Muller (1995), Bekki  \& Chiba (2001), and Bullock \& Johnston (2005).

\subsection{The blue horizontal branch \label{bhb}}

An assessment of the properties of the HB is fundamental in order to properly 
understand the relation existing between globular cluster and field halo
population. At least  three parameters (metallicity, age, and possibly He
abundance) are required to  explain the properties of the HB population in
globular clusters (see Sect.~\ref{hb}). Since  it is not obvious that the
distribution of these parameters is the same for field and cluster stars, it
could be possible to observe systematic differences between their HB
populations. It is difficult to extract clean samples of field HB stars to be
used for this comparison. The best results are obtained for blue HB stars and RR
Lyrae (e.g., Kinman \& Allen 1996), while it is difficult to obtain a complete
sample of red HB stars. Here, we only limit the discussion to the first ones.

\begin{figure}
\centering
\includegraphics[bb=20 160 570 700, clip,scale=0.43]{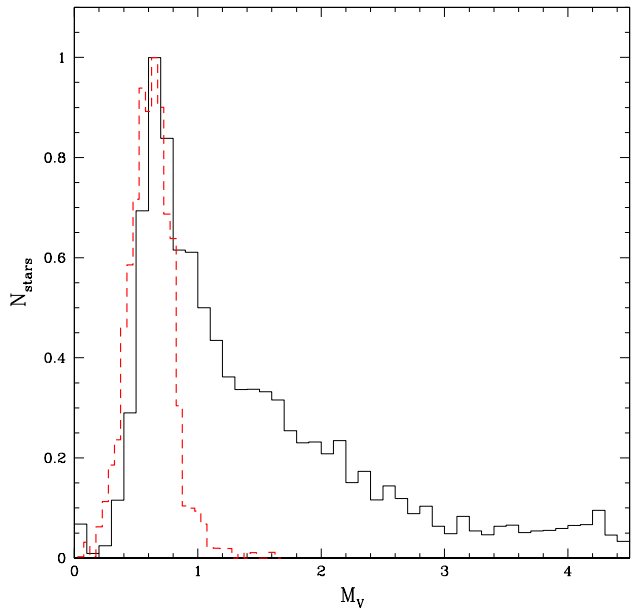}
\caption{Luminosity function for BHB stars in globular clusters (from Piotto et
al. 2003: solid histogram) and in the field (Brown et al. 2008: red dashed
line). The two distributions are normalised to their maximum value.}
\label{f:bhb}
\end{figure}

Reliable samples of BHB stars have been obtained by extensive spectroscopic
surveys (e.g., Beers et al. 2007, and references therein), which provided colour
distribution of BHB stars strongly peaked at colours in the range
$0<(B-V)_0<0.1$\, with very few bluer stars (Kinman \& Allen, 1996; Brown et al.
2008). Brown et al. used these data to infer the luminosity function of field
BHB. In Fig.~\ref{f:bhb} we compare their luminosity function with that for BHB
stars in globular clusters obtained using the HST ``Snapshot Survey" (Piotto et
al. 2003). Roughly half of the cluster BHBs are fainter than $M_V=+1$, while
there is virtually no field BHB star fainter than this magnitude. This confirms
the earlier result by Kinman \& Allen (1996): field and cluster BHB luminosity
functions are different, the first one lacking the extension at faint magnitudes
that has to have an origin strictly tied to the globular cluster environment.

Since there are few Na-rich/O-poor stars in the field, this difference can be
immediately related to the correlation existing between the maximum temperature
of HB stars and the extension of the Na-O anticorrelation (Carretta et al. 2007,
2010a). The maximum temperature of the HB for the field stars at $\log{T_{\rm
eff}}\sim 4.1$\ corresponds to an interquartile of the Na-O distribution
$IQR\leq 0.3$, which is close to the minimum value obtained for any globular
cluster (see Carretta et al. 2009a). If the relation between the blue tails of
the HB and the presence of Na-rich stars is caused by a difference in the He
content (D'Antona et al. 2002; Gratton et al. 2010a; see Sect.~\ref{hb}),
He-rich stars are rare in the field, while they are quite common in globular
clusters. This fact supports the view that field halo stars are related almost
completely to the primordial (Na- and He-poor) population of globular clusters,
that is the first generation.

\subsection{Binaries \label{binaries}}

A significant fraction ($\sim 60$\%) of solar type stars in the solar
neighbourhood are in binary systems (Duquennoy \& Mayor 1991). Wide binaries may
be destroyed in very dense fields (Lada \& Lada 2003). In fact, the outcome of
the interactions between a binary and single stars in a cluster depends
critically on the ratio $x$\ of the binary binding energy to the mean kinetic
energy of cluster stars. In general, binaries having $x>1$\ (hard/close
binaries) will, on average, become more bound as a result of interactions with
single stars; for $x<1$\ (soft/wide binaries), binary/single-star interactions
will generally lead to the disruption of the binary (see e.g., Heggie \& Hut
2003 for a review). The binary frequency and the maximum separation may then be
used to estimate the density of the birth environment (Goodwin 2010). The
frequency of spectroscopic binaries among field metal-poor red giants, for
periods less than 6000 days, is $16\pm 4$\%, which is not significantly
different from that measured for field metal-poor dwarfs ($17\pm 2$\%: Carney et
al. 2003). These values are similar to those obtained for field population I
stars, if we restrict to the same period range.

Binary stars play a crucial role in cluster evolution, both as an energy source 
supporting the cluster's post-core collapse dynamics (see e.g. Gao et al. 1991; 
Goodman \& Hut 1989; Vesperini \& Chernoff 1994; Heggie et al. 2006; Trenti et
al. 2007; Hurley et al. 2007; see also Heggie \& Hut 2003 for a review), and as
potential seeds for the formation of a variety of exotic objects (e.g. LMXBs,
CVs, blue stragglers etc.; see e.g. Ivanova et al. 2006, 2008; Ferraro \&
Lanzoni 2008 and references therein). However, the binary fraction in globular
clusters is certainly much lower than in the field (Romani \& Weinberg 1991). In
a recent survey of photometric binaries deduced from the colour-magnitude
diagram, Milone et al. (2008) estimated that typically only a few per cent of
stars in globular clusters are binaries, with the binary fraction being a strong
function of the cluster mass, ranging from $\sim 2$\% for the most massive
globular clusters, up to $\sim 20$\% for the less massive ones. The mass
weighted average value is low, at most about  one third of that for field halo
stars. 

The low binary fraction in globular cluster is not surprising considering their
high density; the larger fraction of wide binaries among the halo stars
indicates that they did not originate within globular clusters which
subsequently disintegrated, but formed in less dense environments (Ryan 1992).
However, an intriguing result is that the incidence of binaries is very
different among first generation and second generation stars (see D'Orazi et al.
2010a). In these samples (drawn from the outer regions of the globular clusters,
where the fraction of binaries is lower through mass segregation effects, see
Milone et al. 2008), only $\sim 1$\% of second generation stars are in binary
systems, while the fraction of spectroscopic binaries in first generation ($\sim
15$\%) is similar to that observed in field metal-poor stars, i.e., it is
consistent with the close relation we are proposing between first generation and
field halo stars. This result is fully consistent with that obtained by Milone
et al. (2008), if we assume that there are no long period binaries in globular
clusters (which is reasonable), and if we consider that first generation makes
up only one third of the globular cluster population.

The dramatic difference between the frequency of binaries in first generation
and second generation indicates (i) that most of the destruction of binaries (or
lack of formation) among second generation stars occurred before relaxation of
the system, because we do not expect large differences in the destruction rates
between first generation and second generation later; and (ii) that only the
most compact binaries have a significant probability to form or survive in the
very dense regions where second generation stars of globular cluster formed.
Vesperini et al. (2011) were able to model successfully these results in the
hypothesis that second generation stars formed a compact subsystem in the
central regions of a more extended first generation globular cluster as proposed
in different forms in D'Ercole et al. (2008), Carretta et al. (2010), Bekki
(2011), and Caloi \& D'Antona (2011). This close relation between the loose
first generation and the very compact second generation clusters is not obvious
in the Baumgardt et al. (2008) simulations, where a simple power law for 
cluster mass distribution is assumed, but it may be understood if they formed 
within a cooling flow as proposed by D'Ercole et al. (2008). This is a crucial 
ingredient in models of globular cluster formation and evolution, and should
have important consequences on the dynamical models.

Summarizing, we presented several arguments to show how the fraction of the
first generation stars lost by globular clusters may be considered one of main
building block of the halo, although other minor contributors cannot be
excluded. These arguments include their total mass, the metallicity
distribution, the element-to-element abundance ratios, the location within the
Milky Way, the distribution of HB stars, and the binary fraction. If true, this
connection establishes a close relationship between the formation of globular
clusters and that of the halo: a large fraction of the halo stars formed in the
same episodes which ultimately lead to the formation of the globular clusters,
and therefore should have a similar age. On the other hand, the connection is
much looser between metal-rich globular clusters and the thick disk, since the
specific  frequency of globular clusters is reduced by roughly a factor of ten
in this case. This suggests that thick disk stars typically formed in a large
number of smaller episodes of star formation.

\section{Conclusions \label{sec:5}} 

In this review we have presented an update on our knowledge of the multiple
populations in globular clusters. The phenomenon moved from being generally
considered an exotic curiosity, present only in a few, peculiar clusters (such
as e.g., $\omega$~Cen, NGC~2808, M~22), to a feature attributed only to massive
clusters (the ones mentioned above, plus M~54 or NGC~6388), and finally to the
widespread acceptance of its probable universality (at least among clusters
massive enough, such as the Milky Way globular clusters). We cannot fully
understand the intrinsic properties of globular clusters, their formation, and
their interaction with the Milky Way, unless we are able to quantitatively
define the properties of the multiple populations in globular clusters, using a
manifold of information coming from spectroscopy, photometry, and theoretical
modelling. Recent progress in gathering extensive, high quality observational
samples from both spectroscopy and photometry are now allowing to successfully
attack long standing problems such as the second-parameter problem or the
relation between globular clusters and field halo.

Many questions remain still open in understanding how globular cluster formed.
The most critical  concern the relative roles of the different mechanisms that
may cause inhomogeneities in globular clusters, the nature of the polluters
(which sets the timescale) and clarification of all nucleosynthesis mechanisms
involved, the origin of the diluting material, and the construction of reliable
3-d hydrodynamical models. 

While theoretical models have made substantial progress and should continue,
maybe the theory is still lagging behind observations and is unlikely that most
of these questions will be solved without new inputs. Here, still important
progress can be obtained by pursuing the {\it archaeological} approach of
studying old systems that provided most of the evidence available. In
particular, spectroscopy can be used to explore the abundance pattern for those
elements, such as He, Li, Al, and F, whose potential has not been fully
exploited; or to obtain an accurate set of abundances for large sets of stars in
each cluster, in order to study the shape of the distributions (continuous {\it
vs} discrete) or the connection with kinematics or location within the clusters.
Photometry may provide both large samples  of stars and high accuracy for many
more globular clusters. All these studies may likely be conducted with currently
available instrumentation, at least for Milky Way globular clusters, while
possibly new instruments are required to study the extragalactic ones. 

On the other hand, detailed study of star formation histories in large star
forming regions and massive young clusters may also turn out to be very
profitable; thus, it should be intensified and systematically pursued. Most of
the interesting objects are in other galaxies, and require very high quality
observational material, possibly adaptive optics assisted observations and
integral field spectroscopy. Studies of the interstellar matter in these large
star-forming regions with large millimetric arrays such as ALMA might also be
interesting to reveal the same existence of the cooling flows predicted in
globular cluster formation scenarios. 

However, when considering young objects, it should be clear that there must be
something fundamentally different in how formation of massive clusters proceeded
in the {\it halo} and how is going on in {\it disk-like} environments: basic
evidence for this is given by the different specific frequency of globular
cluster in the two environments and the universal presence of clearly separated
blue and red sequences of globular clusters in galaxies (see the review by
Brodie \& Strader 2006). 

Ideally, the connection between globular clusters, their multiple populations,
and the Galactic halo should be explored by directly detecting the streams
formed by the individual globular clusters. This cannot be done at present
because of the short dynamical timescales in the inner Galaxy (a typical inner
halo star will have made between 50 and 100 revolutions in a Hubble time). Each
accreted satellite will then give rise to multiple streams (see Helmi \& White
1999; and discussion in Helmi 2008). Better perspectives are opened by studying
streams in the integral of motion space (Helmi et al. 1999; Kepley et al. 2007).
This requires determining the kinematics of a large number of halo stars. Some 
progress in the field is expected in the near future from the continuing SEGUE
(Beers et al. 2006; Yanny et al. 2009) and RAVE (Steinmetz et al. 2006; Siebert
et al. 2011) surveys. Further data will be produced by spectroscopic surveys
just starting or planned for the immediate future, such as HERMES (Freeman
2011), LAMOST (e.g., Zhao et al. 2006), and the FLAMES public ESO survey (PIs
Gilmore and Randich, see http://www.gaia-eso.eu) that will start in 2012 and
will target more than 10$^5$ stars belonging to all Galactic stellar components.
Finally, a special role will be played by the Gaia satellite (Perryman et al.
2001), due to be launched in mid-2013, with its all-sky information on position,
distance, and proper motion for about 10$^9$ objects, and complete information
(radial velocity, metallicity) for a few million stars (see 
http://www.rssd.esa.int).

\begin{acknowledgements}

We thank Franca D'Antona, Melvyn Davies, Valentina D'Orazi, Sara Lucatello, and
Chris Sneden who thoroughly read and expertly commented the paper. We  thank
Warren Brown for having kindly provided  unpublished material used in the
preparation of Figure~\ref{f:bhb}; Nikos Prantzos, Annibale D'Ercole and Corinne
Charbonnel for sending us material from their unpublished talks. This research
has made use of the SIMBAD database, operated at CDS, Strasbourg, France, and of
NASA's Astrophysics Data System. This work was funded by INAF by the grant INAF
2009 "Formation and Early Evolution of Massive Star Clusters".
\end{acknowledgements}



\end{document}